\documentclass[9pt,twocolumn,twoside]{pnas-new}
\templatetype{pnasresearcharticle} % Choose template 
\setboolean{displaywatermark}{false}
\title{The mixing of polarizations in the acoustic excitations of disordered media with local isotropy}

\author[a,b,1]{Maria Grazia Izzo}
\author[c,b]{Giancarlo Ruocco}
\author[c,d]{Stefano Cazzato}

\affil[a]{Universit\'a degli studi di Roma ''La Sapienza'', Dipartimento di Ingegneria Ingegneria Informatica Automatica e Gestionale Antonio Ruberti, Via Ariosto, 00185 Roma, Italy}
\affil[b]{Istituto Italiano di Tecnologia - Center for Life Nanoscience, Viale Regina Elena, 291
00161 Roma, Italy}
\affil[c]{Universit\'a degli studi di Roma ''La Sapienza'', Dipartimento di Fisica, Piazzale Aldo Moro 5, 00185 Roma, Italy}
\affil[d]{Chalmers University of Technology, Department of Applied Physics, Maskingr\"{a}nd 2, 412 58 Gothenburg, Sweden}
\leadauthor{Izzo} 

\correspondingauthor{\textsuperscript{2}To whom correspondence should be addressed. E-mail: izzo@diag.uniroma1.it}

\begin{abstract}
An approximate solution of the Dyson equation related to a stochastic Helmholtz equation, which {describes} the acoustic dynamics of a three-dimensional isotropic random medium with {elastic} tensor fluctuating in space, is obtained in the framework of the Random Media Theory. The wavevector-dependence of the self-energy is preserved, thus allowing a description of the acoustic dynamics at wavelengths comparable with the size of heterogeneity domains. This in turn permits to quantitatively describe the mixing of longitudinal and transverse dynamics induced by the medium's elastic heterogeneity and occurring at such wavelengths. A functional analysis aimed to attest the mathematical coherence and to define the region of validity in the frequency-wavector plane of the proposed approximate solution is presented, with particular emphasis dedicated to the case of disorder characterized by an exponential decay of the covariance function.
\end{abstract}

\dates{This manuscript was compiled on \today}
\begin{document}

% Optional adjustment to line up main text (after abstract) of first page with line numbers, when using both lineno and twocolumn options.
% You should only change this length when you've finalised the article contents.
\verticaladjustment{-2pt}

\maketitle
\thispagestyle{firststyle}
\ifthenelse{\boolean{shortarticle}}{\ifthenelse{\boolean{singlecolumn}}{\abscontentformatted}{\abscontent}}{}
\newcommand{\sgn}{\text{sgn}}

\section*{Introduction} \label{intro}
Most materials we encounter on a daily basis, such as glasses, polycrystalline aggregates, ceramics, composites, geophysical materials, and concrete can be classified as heterogeneous materials, being composed by domains with different physical characteristics. An acoustic wave propagating in a three-dimensional system can be characterized by its phase velocity, amplitude and polarization. In a heterogeneous medium the acoustic excitations experience retardation, attenuation (Rayleigh anomalies) and depolarization.
{Strong attention has been deserved in literature both to the Rayleigh anomalies and to the mixing of polarizations  \cite{Ferrante,hydro_MonGio,Ruta, Mossa, Baldi, Nakayama, Schirmacher_4th,John,Schi_SCBorn,Shir_gen1, Marruzzo, Matic, Masciovecchio, Ruffle, Ruffle_picosecodspectro, Ruzi, Scopigno1HFglasses_transverse, zanatta_INS_GeO2, Violini_glucose, Cunsolo,  Bencivenga, Cimatoribus}. They have been, however, designed as} disjoint phenomena and never been addressed by an analytical theory as related aspects originating from a common root, the disordered nature of the medium. An analytical theory describing the mixing of polarizations of acoustic excitations in disordered systems is, furthermore, so far lacking. 
One of the challenge in obtaining an unified picture of the above-metioned phenomena is their occurrence on different length-scales. In the so-called Rayleigh region, i.e. for values of wavelength ($\lambda$) of elastic excitations much lower than the characteristic size ($a$) of inhomogeneity domains, the phase velocity of acoustic modes shows a softening with respect to its hydrodynamic value (retardation). It is observed, moreover, a strong increase of the acoustic wave attenuation (Rayleigh scattering), the two quantities being related to each other by Kramers-Kroning relations \cite{Marruzzo}. The coupling between longitudinal and transverse polarizations is instead maximum beyond the Rayleigh region when {$\lambda \sim a$} \cite{Calvet}. 
The basic analytical instrument to describe the ensemble averaged elastodynamic response of a heterogeneous system to an impulsive force is the so-called Dyson equation for the mean field \cite{Sobczyk, B1, Bourret, Turner, Calvet}, introduced in the framework of the Random Media Theory (RMT), or Heterogeneous Elasticity Theory when referring to the specific case of elastic inhomogeneity \cite{Schi_SCBorn, John, Schirmacher_4th, Schirmacher_genteo,Shir_gen1, Marruzzo, Ferrante}. The solution of the Dyson equation, however, can be obtained only under suitable approximations, which {have} a limited wavelengths {range of validity}  \cite{Sobczyk, B1, Bourret}, thus avoiding a unified theoretical description of experimental observations. 

The exact solution of the Dyson equation can be formally cast in a Neumann-Liouville series, the so-called perturbative series expansion. Even if in most real cases a direct sum, or even establishing the criteria of convergence of the series, is not possible, it constitutes the general starting point for smoothing methods or approximations \cite{Sobczyk}. Its truncation to the lowest non-zero order leads to the Born Approximation. We propose an approximate solution of the  vectorial Dyson equation, which takes into account in an approximate form {terms} of the Neumann-Liouville series {up to the second order}, thus introducing corrective terms to the {so-called} Born Approximation. We will refer to it as to a Generalized Born Approximation (GBA). We first derive an analytical expression for the GBA and state the general conditions for its validity in a given region of wavelengths. In a second stage, we analyze the specific case of an exponential decay of the covariance function of the elastic fluctuations. {We} show how in such a case the GBA {can be applied} up to wavelengths of the order of the average size of heterogeneity domains. We then calculate in the GBA frame the current spectra related to acoustic dynamics and show that the GBA allows for a description of all the effects that the topological disorder has on acoustic dynamics, including the Rayleigh anomalies and, for the first time, the mixing between longitudinal and transverse polarization.

The rest of the paper is organized as in the following. In Sec. \ref{Method} we introduce the {GBA}, we discuss its physical significance with the support of the Feynman diagram technique and its relationship with the perturbative series expansion. In Sec. \ref{genborn} we describe with mathematical detail the proposed approximation and demonstrate its validity in a proper domain of the frequency ($\omega$) - wavevector ($q$) plane. In Sec. \ref{domain} we deal with the specific case of an exponential decay of the covariance function and define in this case the domain of validity of the GBA. In Sec. \ref{mixing} we discuss how the GBA can account for the mixing of polarizations. {In Sec. \ref{acoustic_prop} we show what the acoustic dynamics properties accounted by the GBA are, allowing a qualitative comparison with existing experimental results.} Conclusions are outlined in Sec. \ref{conclusion}. Technical details in addition to the main text are reported in two appendices. 
\section{Methods} \label{Method}
\subsection{The Dyson Equation and its approximate solutions in the Random Media Theory} \label{intro2}
The elastic response of an unbounded and elastic medium to an impulsive force can be obtained as a function of the Green's dyadic by solving the so-called stochastic Helmholtz equation \cite{Sobczyk, Turner},
\begin{eqnarray}
\big\{\hat{L}^0_{ki}(\textbf{x},t)+\hat{L}^s_{ki}(\textbf{x},t)\big\}G_{i j}(\textbf{x},\textbf{x}',t)=\delta_{k j}\delta(\textbf{x}-\textbf{x}')\delta(t). \label{stoHelmholtz_1}
\end{eqnarray}
Summation over repeated indices is assumed.
The second-rank Green's dyadic, $G_{ij}(\textbf{x},\textbf{x}',t)$, is the response of the system at the spatial point of vectorial coordinate $\textbf{x}$ in the $i$-th direction at the time $t$ to a unit-impulse at the point $\textbf{x}'$ in the $j$-th direction. The function $\delta_{kj}$ is a Kronecker delta function, $\delta(\textbf{x})$ and $\delta(t)$ are Dirac delta functions. The elastic tensor, $C_{ijkl}(\textbf{x})$, and the density, $\rho(\textbf{x})$, of the system are spatially heterogeneous. We define $C_{ijkl}(\textbf{x})=C_{ijkl}^0+\delta C_{ijkl}(\textbf{x})$, $\rho(\textbf{x})=\rho^0+\delta \rho(\textbf{x})$. We hypothesize statistical homogeneity, thus $C_{ijkl}^0 = <C_{ijkl}(\textbf{x})>$ and $\rho^0 = <\rho(\textbf{x})>$. The brackets $< \ \ >$ denote the ensemble average. The operators $\hat{\textbf{L}}^0(\textbf{x},t)$ and $\hat{\textbf{L}}^s(\textbf{x},t)$ are respectively a deterministic differential operator related to the average, constant in space, elastic tensor and density and a linear stochastic operator accounting for the fluctuating, space-dependent, terms of the same quantities. We assume statistical and local isotropy and express the elastic tensor as a function of the shear modulus, $\mu(\textbf{x}) = \mu_0+\delta \mu(\textbf{x})$, and of the Lam\'e parameter, $\lambda(\textbf{x})= \lambda_0+\delta \lambda(\textbf{x})$. Under these hypotheses the operators $\hat{\textbf{L}}^0(\textbf{x},t)$ and $\hat{\textbf{L}}^s(\textbf{x},t)$ are given by \cite{Turner}
\begin{multline}
\hat{L}^0_{ki}(\textbf{x},t)=-\delta_{ki} \rho_0 \frac{\partial^2}{\partial t^2} +\lambda_0\frac{\partial}{\partial x_{k}}\frac{\partial}{\partial x_{i}}+\mu_0[
\frac{\partial}{\partial x_{k}} \frac{\partial}{\partial x_{i}}+\\ \delta_{ki}\frac{\partial}{\partial x_l}\frac{\partial}{\partial x_l}]; \label{L0} 
\end{multline}
\begin{multline}
\hat{L}^s_{ki}(\textbf{x},t)=-\delta_{ki} \delta \rho(\textbf{x}) \frac{\partial^2}{\partial t^2}+\frac{\partial}{\partial x_{k}}\delta\lambda(\textbf{x})\frac{\partial}{\partial x_{i}}+\\ \frac{\partial}{\partial x_{k}} \delta \mu (\textbf{x})\frac{\partial}{\partial x_{i}}+\delta_{ki}\frac{\partial}{\partial x_l}\delta \mu(\textbf{x})\frac{\partial}{\partial x_l}. \label{Ls}
\end{multline}
We take under exam the case of spatial fluctuations of the elastic tensor, thus the first term in Eq. \ref{Ls} is zero.

The quantity physically relevant, related to the dynamic structure factor, which can be {accessed}, e.g., by Inelastic X-ray Scattering (IXS) or Inelastic Neutron Scattering (INS), is the ensemble averaged Green's dyadic, $<\textbf{G}(\textbf{x},\textbf{x}',t)>$. 
In place of solving the Helmholtz equation (impossible in most cases) and then averaging, one can look for a suitable expression of an effective deterministic operator, $\hat{\textbf{D}}$, such that \cite{Sobczyk}
\begin{eqnarray}
\hat{D}_{ki}(\textbf{x},t) <G_{ij}(\textbf{x},\textbf{x}',t)>=\delta_{kj}\delta(\textbf{x}-\textbf{x}')\delta(t).
\end{eqnarray}
The latter equation is referred to as the Dyson equation. 
Drawbacks in the definition of $\hat{\textbf{D}}$ comes, however, from the fact that the operator $\hat{\textbf{L}}(\textbf{x},t)=\hat{\textbf{L}}^0(\textbf{x},t)+\hat{\textbf{L}}^s(\textbf{x},t)$ cannot be inverted. 

The Dyson equation can be rephrased by setting a formal expression for the average Green's dyadic \cite{Turner},
\begin{multline}
<G_{i j}(\textbf{x},\textbf{x}',t)>= G^0_{i j}(\textbf{x},\textbf{x}',t)+\\ \int\int d\textbf{x}''d\textbf{x}'''G^0_{ik}(\textbf{x},\textbf{x}'',t)\Sigma_{k\alpha}(\textbf{x}'',\textbf{x}''',t)<G_{\alpha j}(\textbf{x}''',\textbf{x}',t)>.\label{Dyson_space}
\end{multline}
The integrals are extended to $\mathbb{R}^3$. $\textbf{G}^0(\textbf{x},\textbf{x}',t)$ is the Green's dyadic of the \lq bare\rq \ medium, solution of the deterministic Helmholtz equation related to the operator $\hat{\textbf{L}}_0(\textbf{x},t)$. Eq. \ref{Dyson_space} is based on the introduction of the so-called mass operator or self-energy, $\mathbf{\Sigma} (\textbf{x},\textbf{x}',t)$, which embeds all the information related to disorder.
The problem of solving the Dyson equation thus translates into finding a suitable expression for the self-energy. The self-energy can be cast in a Neumann-Liouville series by starting from the related stochastic Helmholtz equation, giving rise to the so-called perturbative series expansion. This generally constitutes the starting point for smoothing methods or approximations \cite{Sobczyk, B1}. 

In the Fourier space Eq. \ref{Dyson_space} becomes
\begin{eqnarray}
<\textbf{G}(\textbf{q},\omega)>=\frac{1}{\textbf{G}^0(\textbf{q},\omega)^{-1}-{\mathbf{\Sigma}}(\textbf{q},\omega)}, \label{G_medio}
\end{eqnarray}
where $\textbf{q}$ and $\omega$ are respectively the conjugate variables of $\textbf{x}$ and $t$.
\paragraph{The Born Approximation} \label{born}
Under the hypothesis of statistical homogeneity, truncation of the perturbative series expansion to the lowest non-zero order leads to the so-called Bourret or Born Approximation \cite{Sobczyk,Turner,Calvet, Bourret,B1,B2,Apre,Schi_SCBorn}. In the Fourier space it states
\begin{align}
\Sigma_{k \alpha}^B(\textbf{q},\omega)&=\hat{L}_{1 k \alpha i j}G_{i j}^0(\textbf{q},\omega). \label{Sigma_Bou}
\end{align} 
The operator $\hat{\textbf{L}}_1$ is related to the operator $\hat{\textbf{L}}^s$ defined in Eq. \ref{Ls} by ensemble averaging and Fourier transforming \cite{Turner}. 
Since we only account for fluctuations of the elastic tensor, the operator $\hat{\textbf{L}}_1$ can be expressed by introducing the covariance function of the elastic tensor fluctuations, \textit{old}: $\tilde{R}_{\gamma \alpha j l \beta k i \delta}(\textbf{x}=\textbf{x}_1-\textbf{x}_2)=<\delta C_{\gamma \alpha j l}(\textbf{x}_1)\delta C_{\beta k i \delta}(\textbf{x}_2)>$. Eq. \ref{Sigma_Bou} becomes \cite{Turner}
\begin{multline}
\Sigma^B_{k \alpha}(\textbf{q},\omega)=\hat{L}_{1 k \alpha i j}G_{i j}^0(\textbf{q},\omega) =\\ \int \ d \textbf{q}' \ q_{\beta}q_{l} q'_{\delta} q'_{\gamma} \tilde{R}_{\gamma \alpha j l \beta k i \delta}(\textbf{q}-\textbf{q}')G_{i j}^0(\textbf{q}',\omega), \ \label{Sigma_Bo_F}
\end{multline}
where the wavevector $\textbf{q}'$ is the variable of integration. It is $q=|\textbf{q}|$. The self-energy in the Fourier space can thus be written as a convolution between the \lq bare' Green's dyadic and the Fourier transform of the covariance function of the elastic tensor fluctuations. 
Despite simplicity, the Born Approximation imposes rather strong restrictions both on the intensity of the elastic constants fluctuations and on the values of $q$ and $\omega$ with respect to $a$.
A necessary condition for the validity of the Born Approximation is indeed to deal with small values of the intensity of elastic fluctuations and of wavevector and frequency \cite{B1,B2,Apre}. The condition $\tilde{\epsilon} aq(q_{0i})\ll 1$ shall be met. It is $q_{0i}=\omega/c^0_{i}$, where $c^0_{i}$ is the phase velocity of the acoustic excitations  with $i$-th polarization in the \lq bare' medium. The parameter $\tilde{\epsilon}^2$ is the \lq disorder parameter' \cite{Schirmacher_4th,Marruzzo}, i.e. the square of the intensity of spatial fluctuations of elastic constants normalized to their average value, whereas $\epsilon^2$ represents the same unrenormalized quantity. 
\paragraph{The Self-Consistent Born Approximation} \label{basic_1}
The so-called Self-Consistent Born Approximation (SCBA) \cite{Schirmacher_4th, Schirmacher_genteo, Marruzzo,Ferrante} or Kraichnan model \cite{B2, Kraichnan} can be derived from the more general mean field theory, the Coherent Potential Approximation (CPA) \cite{Soven, Taylor, Elliot, Sheng, Kohler}, under the hypothesis of small fluctuations \cite{Kohler}. It is, however, not affected by the same small wavevectors and frequencies limitation of the Born Approximation \cite{B2,Kohler}. In place of truncating the Neumann-Liouville series {in the SCBA frame} it is constructed an effective nonlinear deterministic equation defining the average Green's dyadic. Because this latter can be related to a realizable model and it can be exactly solved, the SCBA solution will guarantee certain consistency properties \cite{Kraichnan}. 
The self-energy in the SCBA is given by 
\begin{align}
\Sigma_{k \alpha}(\textbf{q},\omega)=&\hat{L}_{1 k \alpha i j}<G_{i j}(\textbf{q},\omega)>. \label{SC}
\end{align} 
Eq. \ref{SC} and the Dyson equation, Eq. \ref{G_medio}, form a set of self-consistent equations. At the step $n=0$ it is $<\textbf{G}(\textbf{q},\omega)>^{n=0}= \textbf{G}^0(\textbf{q},\omega)$. Even if the logic behind the two approaches, i.e. the truncation of the Neumann-Liouville series defining the exact solution of the problem leading to the Born Approximation, or the mean-field approach behind the CPA leading to the SCBA, is different, Eq. \ref{SC} can easily allow a connection between the two: the expression at the first step of the SCBA is the same obtained from the Born approximation.  Accordingly we expect for those cases where it is applicable, the SCBA to provide a better approximation than the Born Approximation. Generalizations of the Born Approximation in the framework of the SCBA have attracted interest in several fields of physics \cite{Jin, Li, Esposito, Igna}.
A link between the SCBA and the perturbative series expansion is discussed in Sec. \ref{basic} by exploiting the Feynman diagram technique. 

An analytical calculation of the self-energy in the SCBA frame is possible by assuming at each step of the self-consistent calculation $q=0$ in the expression of the mass operator \cite{Schirmacher_genteo, Marruzzo, Ferrante,Schirmacher_4th}. 
By exploiting this approach the SCBA revealed to correctly describe the Rayleigh anomalies of acoustic waves in a topologically disordered medium \cite{Marruzzo, Ferrante,Schirmacher_4th}. 
The SCBA thus revealed to give an answer to important questions such as how does the attenuation and phase velocity vary with the wavevectors in the Rayleigh region. We could also expect {that} the SCBA in a three-dimensional space {can} carry information about the polarization properties of the acoustic waves. This kind of study, however, can be hindered by the impossibility to obtain an analytical calculation of the SCBA self-energy for $aq \sim 1$, that is at the edge of the Rayleigh region where the strong acoustic wave attenuation starts to slow down and the mixing of polarizations is expected to get in.
\paragraph{The mixing of polarizations beyond the Born Approximation.} \label{basic_2}

We introduce an approximate method (GBA) for the calculation of $\mathbf{\Sigma}(\textbf{q},\omega)$. It permits to obtain corrective terms to the Born Approximation in the context of the perturbative series expansion, as discussed in the next Section. 
We discuss in Sec. \ref{mixing} how the GBA permits to describe the mixing of polarizations at the boundary of the Rayleigh region ($aq \sim 1$) {and in Sec. \ref{acoustic_prop} how it allows to describe, together with the mixing of polarizations, also the acoustic anomalies occurring in the Rayleigh region ($aq<1$)}. Similar results cannot be achieved by using the Born Approximation, {thus making the two approximations qualitatively different.}

A sharp increase in the attenuation of the acoustic excitations and a related kink in the phase velocity at $aq \sim 1$ are features related to the coupling between longitudinal and transverse dynamics \cite{Calvet}. They can be described by the Born Approximation in the three-dimensional space \cite{Calvet}.
By exploiting the Born Approximation, however, we couldn't unravel the presence of a clear \lq projection' of the transverse into the longitudinal acoustic dynamics, as instead attested by experimental observations in several topologically disordered systems \cite{Ruzi, Scopigno1HFglasses_transverse,zanatta_INS_GeO2,Cunsolo}. With \lq projection' it is meant the occurrence in the longitudinal dynamic structure factor of a peak-like feature centered at frequencies characteristic of the transverse excitations and occurring at sufficiently high wavevectors.
We can attribute such a failure to the fact that the Born Approximation has a limited range of validity in the wavevector space, as discussed above. In particular it shifts towards lower values of wavevector for higher values of the disorder parameter. 
{Depending on the value of the disorder paramater,} its validity at $q \sim a^{-1}$ can thus be questioned.
Since most of the phenomenology observed in real systems, including the Rayleigh anomalies, can however be qualitatively grasped even by the Born Approximation \cite{Calvet} corresponding to first order truncation of the Neumann-Liouville series, we choose to take under consideration the next order approximation in the perturbative series expansion. It corresponds to truncate the SCBA to the second order \cite{B1}. 
It not only permits to obtain a qualitative description of the phenomenology but also to fit experimental outputs for a real system \cite{izzo}. 
In particular, as we discuss in Sec. \ref{mixing}, the GBA permits to describe the \lq projection' of the transverse dynamics observed in the longitudinal dynamics obtaining results which are qualitatively different from what it is possible to achieve with the Born Approximation.

\subsection{Basic considerations}\label{basic}
The physical meaning of the Dyson equation as well as of the related approximations can be better understood with the aid of the Feynman diagram technique \cite{B1}. 
\begin{figure}[h!]
\begin{center}
\includegraphics[scale=0.3]{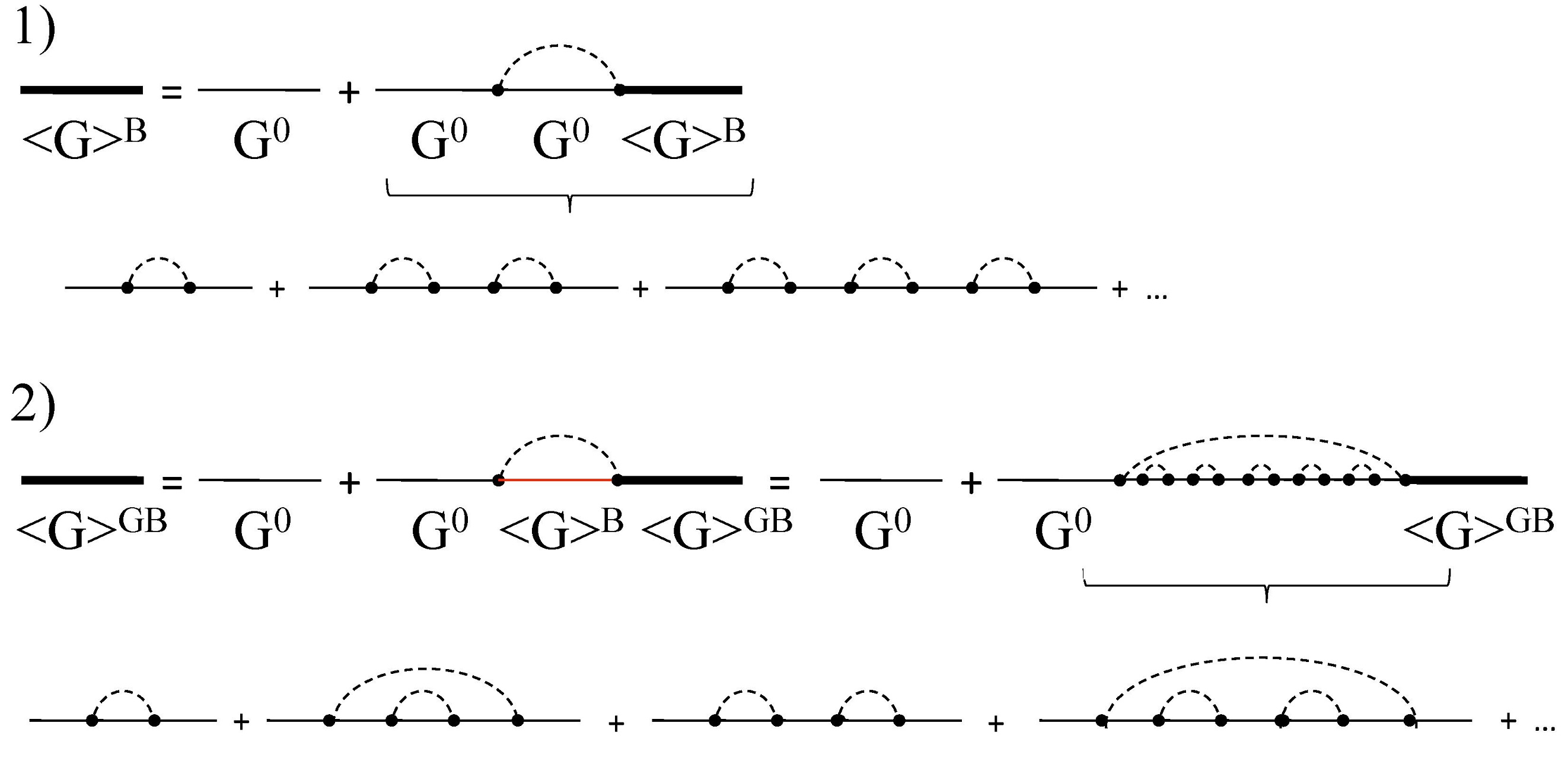}
\caption{\textit{Panel} 1). Feynman diagrams representation of the Dyson equation in the Born Approximation. \textit{Panel} 2).  Feynman diagrams representation of the Dyson equation in the next order approximation of the perturbative series expansion, corresponding to truncation of the SCBA to the second step. It represents the starting point of the GBA.} \label{diagrammi}
\end{center}
\end{figure}
\begin{figure}[h!]
\begin{center}
\includegraphics[scale=0.4]{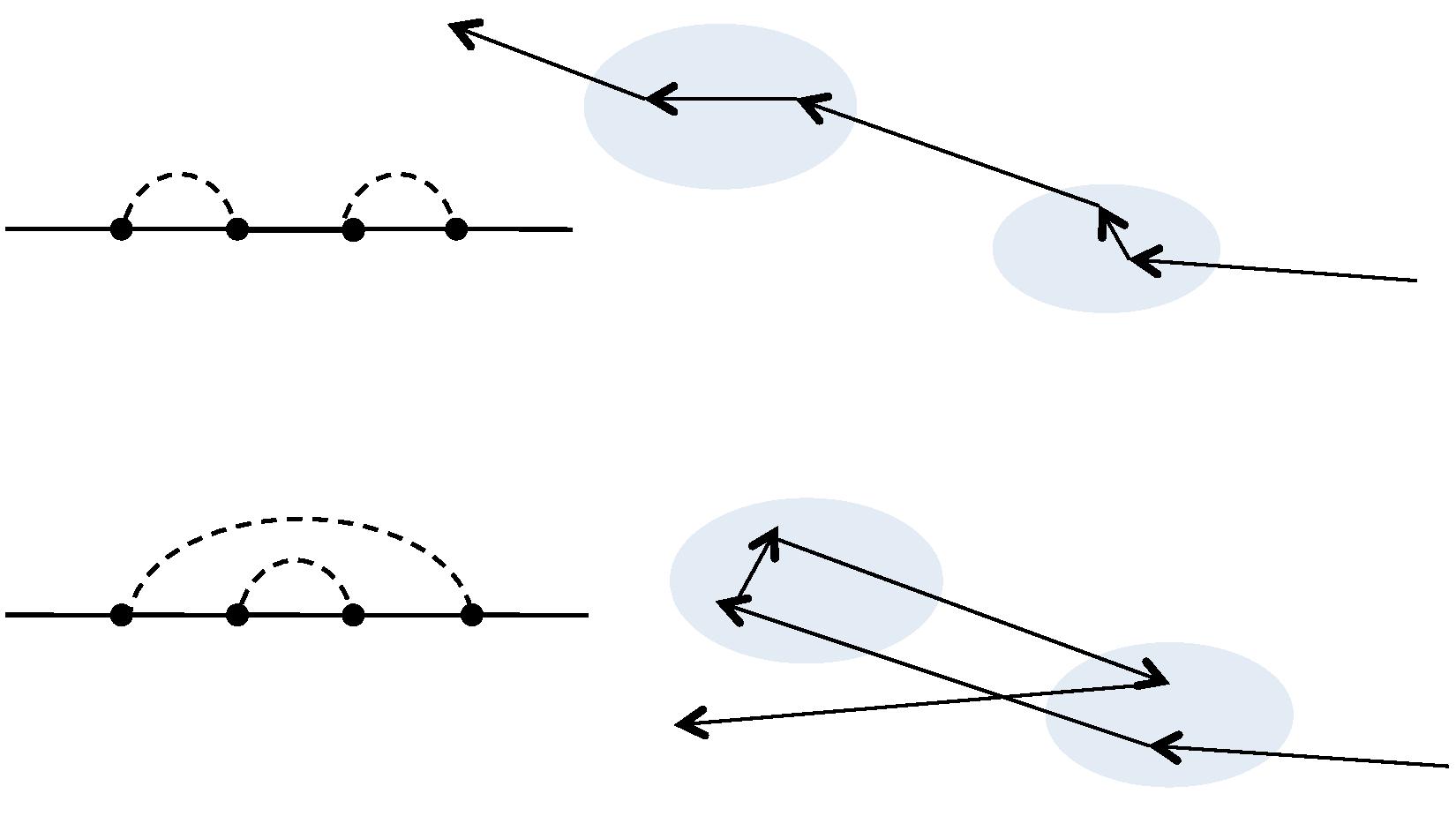}
\caption{Selection of two Feynman diagrams with the corresponding scattering events. Both the diagrams describe a four-fold scattering. The diagram on the top accounts for double scatterings occuring in the same inhomogeneity, whereas  the diagram on the bottom also accounts for a double scattering from two different inhomogeneities.} \label{diagrammi_scat}
\end{center}
\end{figure}
The perturbative series can be rephrased as a sum of appropriate infinite subsequences of the same series. The exact series cannot be summed up, but some of the subsequences can \cite{B1}. Approximations, among which the Born Approximation, are constructed by summation of one or more of the infinite subsequence extracted from the perturbative series \cite{B1}.
It is possible to establish a one to one correspondence between the analytic expressions, which we exploit in this text, and the Faynman diagrams. The diagram technique, however, has the advantage to permit to classify the infinite subsequence entering in the perturbative series depending on scattering events. Within this outlook the Feynman diagrams are classified as strongly or weakly connected diagrams \cite{B1}. Weakly connected diagrams are those that can be always divided into strongly connected diagrams. The self-energy can hence be represented as the {hierarchical} sum of all the strongly connected diagrams. The topology of different strongly connected diagrams is finally related to different kind of multiple scattering events. The Born Approximation, for example, accounts only for double scattering from the same inhomogeneity of an otherwise freely propagating wave, see Figs. \ref{diagrammi} and \ref{diagrammi_scat}. It is indeed obtained through the sum of an infinite subsequence of diagrams, which contains one only {kind of} strongly connected diagram, whose topology describes the above-quoted process. The next order approximation, which will include the next infinite subsequence of diagrams from the exact expansion of the mass-operator, can be obtained with the analytical expression of the mass-operator stated in the Born Approximation (Eq. \ref{Sigma_Bo_F}) by substituting the \lq bare' Green's dyadic, $\textbf{G}^0(\textbf{q},\omega)$, with the approximate expression of the mean Green's dyadic obtained by the Born Approximation itself \cite{B1}. This corresponds to the expression obtained by truncation of the SCBA expression to the second iteration step. In terms of diagram technique, it permits the inclusion of Feynman diagrams accounting for a sequence of scattering between two different inhomogeneities \cite{B1}, see Figs. \ref{diagrammi} and \ref{diagrammi_scat}. Not all possible multiple scattering events are, clearly, included. The approximation presented here can thus be view as a scheme for a partial inclusion of contributions from multiple scattering events. On this {perspective} also the SCBA can be thought as a sum of some of the infinite subsequences composing the perturbative series. Truncation of the SCBA to the third iteration step will include further scattering events not accounted for when the self-energy is obtained by truncation to the second step, and so on. The exact solution of the Dyson equation is unknown. It is thus not possible to establish what is the error related to the SCBA as well as to its truncation to the second step of the iterative procedure. We can however assume that a necessary condition for the truncation to the $n$-th step of the SCBA to give an approximate expression of the self-energy is $|\mathbf{\Sigma}^{n}(\textbf{q},\omega)-\mathbf{\Sigma}^{n-1}(\textbf{q},\omega)|\ll |\mathbf{\Sigma}^{n-1}(\textbf{q},\omega)|$. {On this ground}, Rytov and Kravtsov \cite{B1} established the necessary condition for the validity of the Born approximation previously stated. 
A necessary condition of validity for the proposed approximation can thus be given by the inequality $|\mathbf{\Sigma}^{3}(\textbf{q},\omega)-\mathbf{\Sigma}^{2}(\textbf{q},\omega)|\ll |\mathbf{\Sigma}^{2}(\textbf{q},\omega)|$.
%We discuss in Appendix B when such a condition is verified. 
%
It is shown in Appendix B that in the domain of the $(\omega,q)$ plane where the series representation introduced in Sec. \ref{genborn} approximates the quantity $\mathbf{\Sigma}^2(\textbf{q},\omega)$ this inequality is satisfied if the magnitude of the remainder function of order one {of the series representation} of $\mathbf{\Sigma}^{2}(\textbf{q},\omega)$ is small enough.
It is furthermore shown that in such a domain the necessary condition of validity for the GBA is less stringent than for the Born Approximation.

Depolarization effects in the scattering of electromagnetic waves by an isotropic random medium has been predicted {by exploiting a} second order {representation for the} scattered intensity \cite{Zuniga}. {The scattering of electromagnetic waves by the random media is cast in terms of Green's dyadic and the formal solution of the problem is given in terms of a Neumann iteration series.} The n-th order of the scattered intensity is obtained by truncation of the Neumann series and ensemble averaging. Depolarization effects are also observable even in the first-order scattered intensity from an anisotropic random medium \cite{Zuniga_ani}. We recall that the optical theorem establish a connection between the self-energy and the intensity operator characterizing the Bethe-Salpter equation, which permits to describe the intensity of the mean field \cite{B1}. These results thus emphasize the soundness of our findings. 

The input parameters of the theory are the correlation length, $a$, the disorder parameter, $\tilde{\epsilon}^2$, and the longitudinal and transverse phase velocity of the \lq bare' medium, $c^0_{L(T)}$. 
\section{Results and Discussion}
\subsection{The Generalized Born Approximation} \label{genborn}
Under the hypothesis of local isotropy it is convenient to introduce the orthonormal basis defined by the direction of wave propagation, $\hat{q}$, and the two orthogonal ones \cite{Turner}. On this basis all the \lq bare'  Green's dyadic, average Green's dyadic and self-energy are diagonal.
The \lq bare\rq \ Green's dyadic becomes
\begin{eqnarray}
\textbf{G}^0 (\textbf{q},\omega)=g^0_L(\textbf{q},\omega) \hat{q}\hat{q}+g^0_T(\textbf{q},\omega)(I-\hat{q}\hat{q}), \label{G0_base}
\end{eqnarray}
{with \lq T' and \lq L' labelling transverse and longitudinal modes, respectively.}
The longitudinal and transverse \lq bare\rq \ Green's functions, $g^0_L(\textbf{q},\omega)$ and $g^0_T(\textbf{q},\omega)$ respectively, can be formally written by following a regularization procedure \cite{jack} as
\begin{multline}
g_{L(T)}^0(\textbf{q},\omega)=lim_{\eta \rightarrow 0^+} \frac{1}{(\omega+ic_{L(T)}^0\eta)^2-(c_{L(T)}^0q)^2} = \\ (c_{L(T)}^0)^{-2} p.v.\big\{ \frac{1}{q_{0 L(T)}^2-q^2}\big \}- i\pi(c_{L(T)}^0)^{-2} 
\sgn(q_{0 L(T)}) \cdot \\ \cdot \delta(q_{0 L(T)}^2-q^2), \label{h}
\end{multline}
where $q_{0L(T)}=\frac{\omega}{c_{L(T)}^0}$. In Eq. \ref{h} $\eta$ is a positive real variable, the symbol p.v. states for the Cauchy principal value and $\sgn(x)$ is the sign function of argument $x$. The retarded solution is selected as required by the causality principle \cite{dirac}. 
Furthermore,
\begin{align}
\mathbf{\Sigma}(\textbf{q},\omega)&=\Sigma_L(\textbf{q},\omega)\hat{q}\hat{q}+\Sigma_T(\textbf{q},\omega)(I-\hat{q}\hat{q}); \nonumber \\
<\textbf{G}(\textbf{q},\omega)>&=<g_L(\textbf{q},\omega)>\hat{q}\hat{q}+<g_T(\textbf{q},\omega)>(I-\hat{q}\hat{q}),\label{G_base}
\end{align}
with
\begin{eqnarray}
<g_{L(T)}(\textbf{q},\omega)>=\frac{1}{g_{L(T)}^0(\textbf{q},\omega)^{-1}-\Sigma_{L(T)}(\textbf{q},\omega)}. \label{gL}
\end{eqnarray}
The GBA address an approximate expression of the self-energy obtained by truncation of the perturbative series expansion to the second order. This is obtained by substituting the \lq bare' Green's dyadics in Eq. \ref{Sigma_Bo_F} (Born Approximation) with the  expression of the average Green's dyadic obtained by the Born Approximation \cite{B1}. It is thus equivalent to truncate Eq. \ref{SC} to the second iteration step. We obtain for the diagonal terms of the self-energy,
\begin{multline}
\Sigma_{kk}(\textbf{q},\omega)=  \hat{L}_{1 k k i i}<G_{ii}(\textbf{q},\omega)>^1 =\\ \hat{L}_{1kkii}\frac{1}{\tilde{c}_{i}^2}lim_{\eta \rightarrow 0^+}\big\{\frac{1}{\tilde{q}_{0i, \eta}^2-q^2-\frac{\epsilon^2}{\tilde{c}_{i}^2}q^2\Delta\tilde{\Sigma}_{ii}^1(\textbf{q},\omega_{\eta})}\big\}, 
\label{Sigma_K_ap0}
\end{multline}
where $\tilde{q}_{0i,\eta}=\frac{\omega_{\eta}}{\tilde{c}_i}$, $\omega_{\eta}=\omega+i\tilde{c}_i \eta$, $\tilde{q}_{0i}=\frac{\omega}{\tilde{c}_{i}}$, $\tilde{c}_{i}=[c_{i}^2+\epsilon^2\tilde{\Sigma}^1_{i}(q=0,\omega=0)]^{1/2}$ is the macroscopic velocity of the (first step) perturbed medium, $\Delta \tilde{\Sigma}_{ii}^1(\textbf{q},\omega)=\tilde{\Sigma}_{ii}^1(\textbf{q},\omega)-\tilde{\Sigma}_{ii}^1(0, 0)$ and $\tilde{\Sigma}_{ii}^1(\textbf{q},\omega)=(\epsilon^2q^2)^{-1}\Sigma_{ii}^1(\textbf{q},\omega)$. 
{The suffix $1$ marks a quantity calculated to the first step of the self-consistent procedure. The repeated indexes $kk, ii=L,T$. The longitudinal and transverse self-energy are thus respectively composed by two terms accounting for the coupling with longitudinal and transverse dynamics respectively, i.e. $\Sigma_{L(T)} =\Sigma_{LL(TT)}+\Sigma_{LT(TL)}$. }

The expression in curly bracket in Eq. \ref{Sigma_K_ap0}, $ <G_{ii}(\textbf{q},\omega)>^1$, is then formally expanded in a Taylor series with respect to $\frac{\epsilon^2}{\tilde{c}_{i}^2}q^2\Delta \tilde{\Sigma}^1_{ii}(\textbf{q},\omega_{\eta})$. Theorem I below states that this series is convergent almost everywhere (a.e.) in the domain of the $(\omega,q)$ plane where the conditions $\frac{\epsilon^2}{\tilde{c}_{i}^2}|\Delta \tilde{\Sigma}^1_{ii}(\textbf{q},\omega)| < 1$ and $\mathrm{Im}[\Delta \tilde{\Sigma}^1_{ii}(\textbf{q},\omega)]>0$ are fulfilled. Once identified such a domain, we can then possibly find a sub-domain as specified in Corollary II, where
\begin{multline}
\Sigma_{kk}(\textbf{q},\omega) \sim \\ lim_{\eta \rightarrow 0^+}  \frac{1}{\tilde{c}_{i}^2}\sum_{n=0}^{\infty}\hat{\textbf{L}}_{1k k ii}\{\frac{[\frac{\epsilon^2}{\tilde{c}_{i}^2}q^2\Delta \tilde{\Sigma}_{ii}^{1}(\textbf{q},\omega_{\eta})]^{n}}{[\tilde{q}_{0i,\eta}^2-q^2]^{n+1}}\theta(q_{Max}^{i}-q)\}, \label{approssi}
\end{multline}
$\theta(x)$ being the Heaviside function of argument $x$ and $q^{i}_{Max}$ representing the $q$-boundary of the domain of convergence of the series representation of $<G_{ii}(\textbf{q},\omega)>^1$. Eq. \ref{approssi} provides the expression of the self-energy in the GBA.

The wavevector-dependence of $\Delta \tilde{\Sigma}^1_{ii}(\textbf{q},\omega)$ is determined by the covariance function used to statistically describe the elastic heterogeneity of the system, as established in Eq. \ref{Sigma_Bo_F} above. 
We analyze in detail the case of an exponential decay of the covariance function with correlation length $a$ and amplitude $\epsilon^2$. This choice grounds on simplicity and on the fact that several systems can be described by such a covariance function \cite{Torquato}.
We show in Sec. \ref{domain}, in particular, that in this case the domain of validity of the GBA includes the region $aq \sim 1$, where the mixing of polarization is expected.

\paragraph{Series representation of $<\textbf{G}(\textbf{q},\omega)>^1$.} \label{series}
In the following we demonstrate that it exists a domain of the $(\omega,q)$ plane, where the function $<\textbf{G}(\textbf{q},\omega)>^1$ admits a.e. the power series expansion specified in the following Theorem I.

\medskip
\textbf{Theorem I}. 
\textit{If, being $q, \omega \in \mathbb{R}$, \begin{enumerate}
\item [\textit{i})] $|\Delta \tilde{\Sigma}_{ii}^1(\textbf{q},\omega)| \in {C}^0$;
\item [\textit{ii})] $|\Delta \tilde{\Sigma}_{ii}^1(\textbf{q},\omega)| \leq \Delta\tilde{\Sigma}^{1,Max}_{i}(\omega)$, $q \in [0,q_{Max}^{i}]$, {eventually $q_{Max}^{i} \rightarrow \infty$};
\item [ \textit{iii})] $\mathrm{Im} [\Delta \tilde{\Sigma}_{ii}^1(\textbf{q},\omega)]>0$, $\forall q, \omega \neq 0$;
\end{enumerate} for $q_{Max}^{i}$, $\omega$ and $\epsilon^2$: $\omega \neq 0$, $\frac{\epsilon^2}{\tilde{c}_{i}^2}\Delta \tilde{\Sigma}_{i}^{1,Max}(\omega)<1$,}
\textit{the series $lim_{\eta \rightarrow 0^+} \sum_{n=0}^{\infty} \frac{[\frac{\epsilon^2}{\tilde{c}_{i}^2}q^2\Delta \tilde{\Sigma}_{ii}^{1}(\textbf{q},\omega_{\eta})]^n}{(\tilde{q}_{0 i, \eta}^2-q^2)^{n+1}}$ converges a. e. in the q-interval $[0,q_{Max}^{i}]$ to the function $lim_{\eta \rightarrow 0^+} \frac{1}{\tilde{q}_{0 i, \eta}^2-q^2-\frac{\epsilon^2}{\tilde{c}_{i}^2}q^2\Delta \tilde{\Sigma}_{ii}^{1}(\textbf{q},\omega_{\eta})}$.}

Before to proceede with the proof, we observe that Theorem I can be applied to cases easily realizable by real systems. Because $\mathrm{Im}[\Delta \tilde{\Sigma}_{ii}^1(\textbf{q},\omega)]$ is proportional to the attenuation of the acoustic excitations in the random medium calculated to the first order of the perturbative series expansion, in point \textit{iii)} it is required that such an attenuation is finite for finite values of $q$ and $\omega$. {Furthermore, the series a. e. convergence is ensured} in a region of wavevectors {where} the acoustic excitations in the random medium can still be described through a finite and sufficiently small correction with respect to the acoustic excitations in the \lq bare' medium, {being there the self-energy sufficiently small}.

From the algebraic equality $\frac{1}{A-B}=\frac{1}{A}+\frac{1}{A}\frac{B}{A-B}$, it follows that
\begin{multline}
lim_{\eta \rightarrow 0^+} \int_{0}^{q_{Max}^{i}}dq \frac{1}{\tilde{q}_{0i,\eta}^2-q^2-\frac{\epsilon^2}{\tilde{c}_{i}^2}q^2\Delta \tilde{\Sigma}_{ii}^{1}(\textbf{q},\omega_{\eta})} =\\ \sum_{n=0}^{N}lim_{\eta \rightarrow 0^+} \int_{0}^{q_{Max}^{i}}dq  \frac{[\frac{\epsilon^2}{\tilde{c}_{i}^2}q^2\Delta \tilde{\Sigma}_{ii}^{1}(\textbf{q},\omega_{\eta})]^n}{(\tilde{q}_{0i,\eta}^2-q^2)^{n+1}}+R_N^{i}. \nonumber 
\end{multline}
The remainder function is defined as
\begin{multline}
R_N^{i} = lim_{\eta \rightarrow 0^+} \\ \int_{0}^{q_{Max}^{i}}dq \ \frac{[\frac{\epsilon^2}{\tilde{c}_{i}^2}q^2\Delta \tilde{\Sigma}_{ii}^{1}(\textbf{q},\omega_{\eta})]^N}{(\tilde{q}_{0 i,\eta}^2-q^2)^{N+1}} \frac{\frac{\epsilon^2}{\tilde{c}_{i}^2}q^2\Delta \tilde{\Sigma}_{ii}^{1}(\textbf{q},\omega_{\eta})}{\tilde{q}_{0 i,\eta}^2-q^2-\frac{\epsilon^2}{\tilde{c}_i^2}q^2\Delta \tilde{\Sigma}_{ii}^{1}(\textbf{q},\omega_{\eta})}. 
\nonumber
\end{multline}
We demonstrate in the following that $|R_N^i|$ admits an upper bound, i.e.
\begin{multline}
|R_N^{i}|\leq M_{i} lim_{\eta \rightarrow 0^+}\int_{0}^{q_{Max}^{i}}dq \frac{[\frac{\epsilon^2}{\tilde{c}_{i}^2}q^2\Delta \tilde{\Sigma}_{i}^{1, Max}(\omega)]^N}{|\tilde{q}_{0i,\eta}^2-q^2|^{N+1}} \leq \\ M_{i} [\frac{\epsilon^2}{\tilde{c}_{i}^2}\Delta \tilde{\Sigma}_{i}^{1, Max}(\omega)]^N \frac{N+1}{2N+1} \frac{\pi}{\tilde{q}_{0 i}}, 
\label{serie_resto}
\end{multline}
where $M_{i}=\sup_{q\in[0,q_{Max}^{i}]} |\frac{\frac{\epsilon^2}{\tilde{c}_{i}^2}q^2\Delta \tilde{\Sigma}_{ii}^{1}(\textbf{q},\omega)}{\tilde{q}_{0i}^2-q^2-\frac{\epsilon^2}{\tilde{c}^2}q^2\Delta \tilde{\Sigma}_{ii}^{1}(\textbf{q},\omega)}|$. The latter quantity exists as a consequence of the hypotheses of Theorem I. It is indeed immediate to recognize that for $q, \omega \in \mathbb{R}$ and $\omega \neq 0$, if $\mathrm{Im} [\Delta \tilde{\Sigma}_{i i}^1(\textbf{q},\omega)]$ is strictly positive, the function $|\frac{1}{\tilde{q}_{0i}^2-q^2-\frac{\epsilon^2}{\tilde{c}_{i}^2}q^2\Delta \tilde{\Sigma}_{ii}(\textbf{q},\omega)}|$ is bounded, do not having poles.

To prove the inequality in the third side of Eq. \ref{serie_resto} we need to show that
\begin{eqnarray}
&lim_{\eta \rightarrow 0^+}\int_{0}^{q_{Max}^{i}}dq \frac{[q^2]^N}{|(\tilde{q}_{0i}+i\eta)^2-q^2|^{N+1}}\leq \frac{(N+1)}{(2N+1)} \frac{\pi}{\tilde{q}_{0i}}, \label{dis}
\end{eqnarray} 
where for sake of simplicity we renamed the variable $\frac{\eta}{\tilde{c}_{i}}$ as $\eta$. 
We discuss only the case $\tilde{q}_{0 i}<q_{Max}^{i}$. The case $\tilde{q}_{0 i}>q_{Max}^{i}$ can be easily reconducted to the former by noticing that $\int_{0}^{q_{Max}^{i}}dq \frac{[q^2]^N}{|(\tilde{q}_{0i}+i\eta)^2-q^2|^{N+1}}\leq \int_{0}^{\infty}dq \frac{[q^2]^N}{|(\tilde{q}_{0i}+i\eta)^2-q^2|^{N+1}}$.

We observe that
\begin{eqnarray}
&\frac{1}{2}[\frac{1}{z^N}+\frac{1}{\overline{z}^N}]=\frac{\cos(N\theta)}{|z|^N}; 
\frac{1}{2}[\frac{1}{z^N}-\frac{1}{\overline{z}^N}]=-i\frac{\sin(N\theta)}{|z|^N},
\label{int_ex_1}
\end{eqnarray}
where $z$ is a generic complex variable, $\overline{z}$ is its complex conjugate and $\theta=\arg(z)$.
Furthermore
\begin{multline}
\frac{1}{|z|^N}\leq \frac{|\cos(N\theta)|}{|z|^N}+\frac{|\sin(N\theta)|}{|z|^N}= \frac{1}{2} \sgn\{\cos(N\theta)\} \\ [\frac{1}{z^N}+ \frac{1}{\overline{z}^N}] 
+ \frac{1}{2} i\ \ \sgn\{\sin(N\theta)\} [\frac{1}{z^N}-\frac{1}{\overline{z}^N}]. 
\label{int_ex_2}
\end{multline}
The inequality $1=[\sin^2(x)+\cos^2(x)]^{\frac{1}{2}}\leq |\sin(x)|+|\cos(x)|$ has been exploited. We furthermore considered that $|cos(x)|=cos(x) \cdot \sgn\{ cos(x)\}$. The same applies to $\sin(x)$.
It is thus
\begin{widetext}
\begin{eqnarray}
&lim_{\eta \rightarrow 0^+}\int_{0}^{q_{Max}^{i}}dq \frac{[q^2]^N}{|q^2-(\tilde{q}_{0i}+i\eta)^2|^{N+1}} \leq \frac{1}{2}lim_{\eta \rightarrow 0^+} \int_{0}^{q_{Max}^{i}} dq \Big[\frac{[q^2]^N}{[q^2-(\tilde{q}_{0i}+i\eta)^2]^{N+1}}+\frac{[q^2]^N}{[q^2-(\tilde{q}_{0i}-i\eta)^2]^{N+1}}\Big] \sgn\{\cos[(N+1)\theta(\eta)]\}+ \nonumber \\&+\frac{1}{2} i lim_{\eta \rightarrow 0^+}\int_{0}^{q_{Max}^{i}} dq \Big[\frac{[q^2]^N}{[q^2-(\tilde{q}_{0i}+i\eta)^2]^{N+1}}- \frac{[q^2]^N}{[q^2-(\tilde{q}_{0i}-i\eta)^2]^{N+1}}\Big] \sgn\{\sin[(N+1)\theta(\eta)]\}, 
\label{int_ex_3_0}
\end{eqnarray}
\end{widetext}
where $\theta(\eta)=\arg\{q^2-(\tilde{q}_{0i}+i\eta)^2\}$.
In the framework of a generalization of the Sokhotski-Plemelj theorem \cite{Ple} due to Fox \cite{Fox} it is possible to show that \cite{Galapon}
\begin{eqnarray}
&lim_{\eta \rightarrow 0^+} \int_{a}^{b} dx \frac{f(x)}{[x-(x_0\mp i\eta)]^{N+1}}=\int_{\gamma^{\pm}(x_0)} dz \frac{f(z)}{(z-x_0)^{N+1}}, 
\label{int_ex}
\end{eqnarray}
where $a$, $b$, $x_0$ and $x$ are real variables: $a<x_0<b$, $f(x)$ is a function which admits a complex extension $f(z)$ that is analytic in a region of the complex plane containing the interval $[a,b]$ but not $x_0$, $R_{x_0}=R \backslash \{ x_0\}$, $\gamma^{\pm}(x_0)$ is a path of the region $R_{x_0}$ from $a$ to $b$ belonging to the upper (lower) half-plane of the complex plane.
The second side of Eq. \ref{int_ex_3_0} can thus be rephrased as
\begin{widetext}
\begin{multline}
\frac{1}{2} \Big(\int_{\gamma^-(\tilde{q}_{0i})} \sgn\{\cos[(N+1)\theta]\} + \int_{\gamma^+(\tilde{q}_{0i})}\sgn\{\cos[(N+1)\overline{\theta}]\}\Big) \frac{[z^2]^N}{(z^2-\tilde{q}_{0i}^2)^{N+1}} \ dz+ \frac{1}{2}i \Big(\int_{\gamma^-(\tilde{q}_{0i})} \sgn\{\sin[(N+1)\theta]\} - \\ \int_{\gamma^+(\tilde{q}_{0i})} \sgn\{\sin[(N+1)\overline{\theta}]\}\Big)  \frac{[z^2]^N}{(z^2-\tilde{q}_{0i}^2)^{N+1}} \ dz= \# \int_{0}^{q_{Max}^{i}} dq\frac{[q^2]^N}{(q^2-\tilde{q}_{0i}^2)^{N+1}}+\pi\frac{1}{N!}\frac{d^N}{dz^N}\frac{[z^2]^N}{(z+\tilde{q}_{0i})^{N+1}}|_{z=\tilde{q}_{0i}}, 
\label{int_ex_3_1}
\end{multline}
\end{widetext}
where $\gamma^{\pm}(\tilde{q}_{0i})$ is the contour of the upper (lower) complex half-plane obtained by deformation of the segment $[0,q_{Max}^{i}]$ around $\tilde{q}_{0i}$ by an infinitesimal arc of circle of radius $\phi$ passing around $\tilde{q}_{0i}$ clockwise (counterclockwise). Furthermore it is $\theta=Arg(z^2-\tilde{q}_{0i}^2)$ and $\overline{\theta}=Arg(\overline{z}^2-\tilde{q}_{0i}^2)$. The symbol $\#$ denotes the Hadamard Finite-Part Integrals (or Cauchy Principal Value when $N=0$) \cite{Ple,Fox,Galapon}. We observe that i) $\overline{\theta}=-\theta$; ii) for $z \in \mathbb{R}$ it is $\theta =\overline{\theta}=0$; iii) for $z \in \mathbb{C}$ it is $\sgn\{\sin[(N+1)\overline{\theta}]\}|_{z \in \gamma^+(\tilde{q}_{0i})}=\sgn\{\sin[(N+1)\theta]\}|_{z \in \gamma^-(\tilde{q}_{0i})}$, $\sgn\{\cos[(N+1)\overline{\theta}]\}|_{z \in \gamma^+(\tilde{q}_{0i})}=\sgn\{\cos[(N+1)\theta]\}|_{z \in \gamma^-(\tilde{q}_{0i})}$. The last passage in Eq. \ref{int_ex_3_1} follows from i) the fact that \cite{Galapon}
\begin{multline}
\frac{1}{2}\big(\int_{\gamma^+(x_0)} dz \frac{f(z)}{(z-x_0)^{N+1}} +\int_{\gamma^-(x_0)} dz \frac{f(z)}{(z-x_0)^{N+1}}\big)=\\ \# \int_{a}^{b} dx \frac{f(x)}{(x-x_0)^{N+1}}, 
\label{int_ex_4}
\end{multline}
and ii) from the Residue Theorem, 
\begin{multline}
\int_{\gamma^-(x_0)} \frac{f(z)}{(z-x_0)^{N+1}}-\int_{\gamma^+(x_0)} \frac{f(z)}{(z-x_0)^{N+1}} =\\ 2\pi i Res^{(N+1)}[f(z),x_0]= 2\pi i \frac{1}{N!} \frac{d^N}{dz^N}[f(z)(z-x_0)^{N+1}]|_{z=x_0},
\label{int_ex_4}
\end{multline}
where $Res^{(N)}[f(z),x_0]$ is the residue of order $N$ of the function $f(z)$ around the pole at $z=x_0$ enclosed in the closed path of the complex plane $\gamma^{-}(x_0)-\gamma^{+}(x_0)$. 
It is 
\begin{eqnarray}
\# \int_{0}^{q_{Max}^{i}} dq\frac{[q^2]^N}{(q^2-\tilde{q}_{0i}^2)^{N+1}} \leq  \# \int_{0}^{\infty} dq\frac{[q^2]^N}{(q^2-\tilde{q}_{0i}^2)^{N+1}}.
\end{eqnarray}
Using integration by parts and the Stokes' formula one obtains \cite{Quian}
\begin{multline}
\# \int_{0}^{\infty} dq\frac{[q^2]^N}{(q^2-\tilde{q}_{0i}^2)^{N+1}}=\\ \ p.v. \int_{0}^{\infty} dq \frac{1}{q-\tilde{q}_{0i}}\frac{1}{N!} \frac{d^N}{dq^N}[\frac{[q^2]^N}{(q+\tilde{q}_{0i})^{N+1}}]. \label{HFP}
\end{multline}
The latter equality ensures the existence of the Hadamard Finite-Part Integral, because the Cauchy p.v. exists as a consequence the fact that the $N$-th order derivative of the function $\frac{[q^2]^N}{(q+\tilde{q}_{0i})^{N+1}}$. It is possible to exploit the Residue Theorem to calculate the integral in Eq. \ref{HFP} and verify that such an integral is equal to zero. 
In Appendix B we furthermore prove that
\begin{eqnarray}
& \frac{1}{N!} \frac{d^N}{dz^N}[f(z)(z-x_0)^{N+1}]|_{z=x_0}=\frac{N+1}{2N+1} \frac{1}{\tilde{q}_{0 i}},
\end{eqnarray}
The validity of Eq. \ref{dis} hence follows from Eq. \ref{int_ex_3_1}. 

From Eq. \ref{serie_resto} we can finally conclude that if $\frac{\epsilon^2}{\tilde{c}_{i}^2}\Delta \tilde{\Sigma}^{1,Max}_{i}(\omega)<1$ and $\omega \neq 0$, it is $lim_{N\rightarrow \infty} |R_N^{i}|=0$.
Under these conditions we proved that
\begin{multline}
lim_{\eta \rightarrow 0^+} \int_{0}^{q_{Max}^{i}}dq \frac{1}{\tilde{q}_{0i,\eta}^2-q^2-\frac{\epsilon^2}{\tilde{c}_{i}^2}q^2\Delta \tilde{\Sigma}_{i i}^{1}(\textbf{q},\omega_{\eta})}=\\ \sum_{n=0}^{\infty}lim_{\eta \rightarrow 0^+}  \int_{0}^{q_{Max}^{i}}dq \frac{[\frac{\epsilon^2}{\tilde{c}_{i}^2}q^2\Delta \tilde{\Sigma}_{i i}^{1}(\textbf{q},\omega_{\eta})]^n}{(\tilde{q}_{0i,\eta}^2-q^2)^{n+1}},
\label{serie_resto2}
\end{multline}
thus finally proving \cite{Kolmogorov} Theorem I .

\paragraph{Validity of the Generalized Born Approximation} \label{validity}

{It is worth at this point to provide the expression for the operator $\hat{\textbf{L}}_1$, starting from Eq. \ref{Sigma_Bo_F}, under the hypothesis of local isotropy in the orthonormal basis defined above.} 
{By performing appropriate inner product on the tensor $\tilde{\textbf{R}}(\textbf{q})$, it is obtained \cite{Turner},
\begin{multline}
\Sigma_{kk}(\textbf{q},\omega)=\hat{\textbf{L}}_{1k k ii}<G_{ii}(\textbf{q},\omega)>^1 =\\ \epsilon^2 q^2 \int d \textbf{q}' c(|\textbf{q}-\textbf{q}'|) L_{kk ii}(\hat{qq'})  <G_{ii}(\textbf{q}',\omega)>^1,\label{Sig}
\end{multline}
where $c(\textbf{q})$ is the scalar covariance function of the elastic constants fluctuations, real and positive-defined \cite{Torquato}; $L_{kk ii}(\hat{qq'})$ is a function of the angle $\hat{qq'}$ between the two versors $\hat{q}$ and $\hat{q}'$, resulting from the inner product \cite{Turner, Calvet} also accounting for the transverse degeneracy. The assumption of isotropy allows this function to depend only on the angle $\hat{qq'}$. By making use of spherical coordinates we finally achieve
\begin{multline}
\Sigma_{kk}(\textbf{q},\omega)= \epsilon^2 q^2 \tilde{\Sigma}_{kk}(\textbf{q},\omega)=\\ \epsilon^2 q^2 \int_{-1}^1dx L_{kkii}(x) 2 \pi \int_{0}^{\infty} dq' \ q'^2 c(q,q',x)  <G_{ii}(\textbf{q}',\omega)>^1, \ \ \label{Sig_1}
\end{multline}
with $x=cos(\hat{qq'})$. The function $\tilde{\Sigma}_{kk}(\textbf{q},\omega)$ is implicitly defined in Eq. \ref{Sig_1}.}

The validity of the Generalized Born Approximation follows from the following two corollaries of Theorem I.

\medskip

\textbf{Corollary I}.
\textit{If the covariance function $c(q,q',x) \in C^0$ for $q' \in [0,q_{Max}^{i}]$, then
\begin{eqnarray}
& lim_{\eta \rightarrow 0^+} \int_{0}^{q_{Max}^{i}}dq' \ q'^2 c(q,q',x) \frac{1}{\tilde{q}_{0i,\eta}^2-q'^2-\frac{\epsilon^2}{\tilde{c}_{i}^2}q'^2\Delta \tilde{\Sigma}_{i i}^{1}(\textbf{q}',\omega_{\eta})} = \nonumber \\&=\sum_{n=0}^{\infty} lim_{\eta \rightarrow 0^+}  \int_{0}^{q_{Max}^{i}}dq' \ q'^2 c(q,q',x) \frac{[\frac{\epsilon^2}{\tilde{c}_{i}^2}q'^2\Delta \tilde{\Sigma}_{i i}^{1}(\textbf{q}',\omega_{\eta})]^n}{(\tilde{q}_{0i,\eta}^2-q'^2)^{n+1}}. 
\nonumber
\end{eqnarray}}
The function $q'^2c(q,q',x)$ for $x\in [-1,1]$ and $q'\in [0,q_{Max}^{i}]$ is continuous and bounded. Corollary I thus follows immediately from Theorem I.

We recast the function $\tilde{\Sigma}_{kk}(\textbf{q},\omega)$ in Eq. \ref{Sig_1} as
\begin{multline}
\tilde{\Sigma}_{kk}(\textbf{q},\omega)= \int_{-1}^1dxL_{kk i i}(x)\frac{2\pi}{\tilde{c}_{i}^2} \cdot \\ \cdot \int_{0}^{q_{Max}^{i}} dq' \ q'^2c(q,q',x) \frac{1}{\tilde{q}_{0i}^2-q'^2-\frac{\epsilon^2}{\tilde{c}_{i}^2}q'^2 \Delta \tilde{\Sigma}_{i i}^1(\textbf{q}',\omega)}+\tilde{R}_{k}(\textbf{q},\omega, \epsilon^2), 
\label{resto_magqM_1_3}
\end{multline}
where the remainder function, $\tilde{R}_k(\textbf{q},\omega, \epsilon^2)$, is
\begin{multline}
\tilde{R}_{k}(\textbf{q},\omega,\epsilon^2) = \int_{-1}^1dxL_{kk i i}(x)\frac{2\pi}{\tilde{c}_{i}^2}\cdot \\ \cdot \int_{q_{Max}^{i}}^{\infty} dq' \ q'^2c(q,q',x) \frac{1}{\tilde{q}_{0i}^2-q'^2-\frac{\epsilon^2}{\tilde{c}_{i}^2}q'^2 \Delta \tilde{\Sigma}_{i i}^1(\textbf{q}',\omega)}.
\label{I_decq}
\end{multline} 

\textbf{Corollary II}. 
\textit{For those values of $\epsilon^2$, $q$ and $\omega$: $| \tilde{R}_{k}(\textbf{q},\omega,\epsilon^2)|\ll 1$, it is 
\begin{multline}
\tilde{\Sigma}_{kk}(\textbf{q},\omega) \sim \sum_{n=0}^{\infty} F_k^n(\textbf{q},\omega) = \sum_{n=0}^{\infty}lim_{\eta \rightarrow 0^+} \int_{-1}^1dxL_{kk i i}(x) \frac{2 \pi}{\tilde{c}_{i}^2} \cdot \\ \cdot \int_{0}^{q_{Max}^{i}}dq' \ q'^2c(q,q',x) \frac{[\frac{\epsilon^2}{\tilde{c}_{i}^2}q'^2\Delta \tilde{\Sigma}_{i i}^{1}(\textbf{q}',\omega)]^n}{(\tilde{q}_{0i,\eta}^2-q'^2)^{n+1}}. \label{serieCo}
\end{multline}}
The generic term of the series, $F_k^n(\textbf{q},\omega)$, is implicitly defined in Eq. \ref{serieCo}. 
Corollary II follows from Corollary I. We emphasize, furthermore, that {the validity of} of Corollary II {is constraint to the assumption of} negligible contribution of the self-energy to the average Green's dyadic if $\frac{\epsilon^2}{\tilde{c}_i^2}\frac{|\Sigma_{kk}(\textbf{q},\omega)|}{q^2}\ll 1$. Corollary II is, however, still valid when this hypothesis is violated but it is possible to show that $| \tilde{R}_{k}(\textbf{q},\omega,\epsilon^2) |\ll |\int_{-1}^1dxL_{kk i i}(x)\frac{2 \pi}{\tilde{c}_{i}^2} \int_{0}^{q_{Max}^{i}} dq' \ q'^2c(q,q',x) \frac{1}{\tilde{q}_{0i}^2-q'^2-\frac{\epsilon^2}{\tilde{c}_{i}^2}q'^2 \Delta \tilde{\Sigma}_{i i}^1(\textbf{q}',\omega)}|$.
\subsection{The case of an exponential decay of the covariance function} \label{domain}
We analyze in detail the case of a covariance function cast in the form of an exponential decay function, {finding} the domain of validity of the GBA {in the $(\omega, q)$ plane}.
We {furthermore} consider only spatial fluctuations of the shear modulus. The results, however, can be easily generalized by including {also} spatial fluctuations of the Lam\'e parameter \cite{Turner}. 

In a three-dimensional Fourier space the covariance function in this case reads 
\begin{equation}
c(\textbf{q})= \frac{1}{\pi^2}\frac{q^2a^{-1}}{(q^2+a^{-2})^2}\label{mu},
\end{equation}
where $\int d^3q \ c(\textbf{q})=1$. {Furthermore it is} $\epsilon^2=\tilde{\epsilon}^2\mu_0^2=\delta \mu^2$, being $\delta \mu $ the intensity of the spatial fluctuations of the shear modulus per density. From Eq. \ref{mu} we obtain that in Eq. \ref{Sig_1} it is $c(q,q',x)=\frac{a}{\pi^2}\frac{(aq')^2}{(1+(aq')^2+(aq)^2-2(aq')(aq)x)^2}$. 

We {first} verify in the following that the hypotheses of Theorem I are verified when the covariance function {is given by} Eq. \ref{mu}. The validity of Corollary I follows immediately from the continuity of the function in Eq. \ref{mu}. It is finally possible to find a domain of the $(\omega,q)$ plane where Corollary II {holds}. In this domain the GBA can be {exploited}. It covers a $q$-range up to $\sim a^{-1}$. 
\begin{figure*}[h!]
\begin{center}
\includegraphics[scale=1]{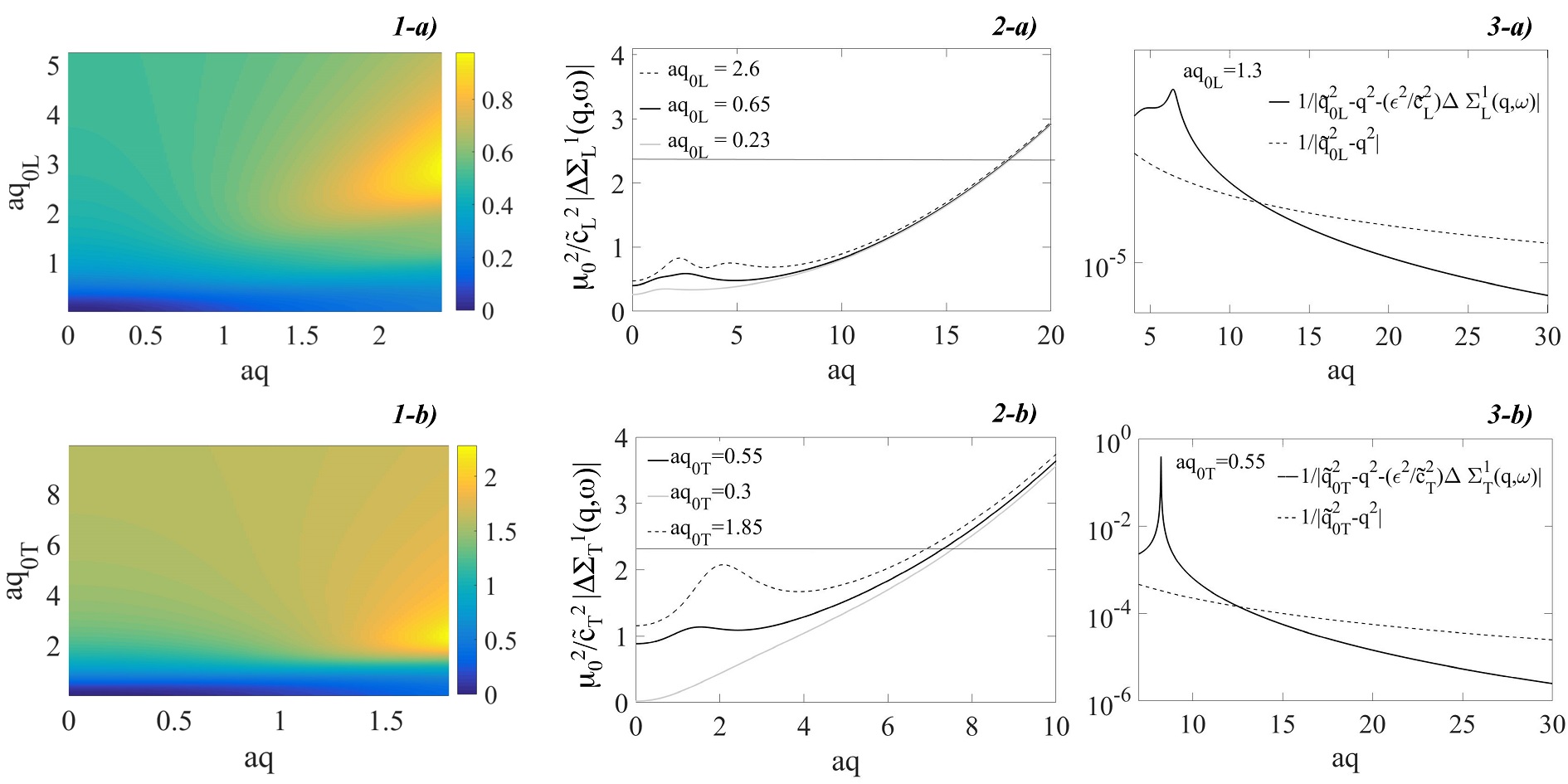}
\caption{\textit{Panels} 1. Projection on the $(\omega,q)$ plane of $a)$ $\frac{\mu_0^2}{\tilde{c}_L^2}|\Delta \tilde{\Sigma}_{L}^1(\textbf{q},\omega)|$ and $b)$ $\frac{\mu_0^2}{\tilde{c}_T^2}|\Delta \tilde{\Sigma}_{T}^1(\textbf{q},\omega)|$. \textit{Panels} 2. $\frac{\mu_0^2}{\tilde{c}_{L(T)}^2}|\Delta \tilde{\Sigma}_{L(T)}^1(\textbf{q},\omega)|$ as a function of $aq$ for three selected values of  $aq_{0L(T)}$. The straight line fixes the value of $(\tilde{\epsilon}^2)^{-1}$. \textit{Panels} 3. $\frac{1}{|\tilde{q}_{0L(T)}^2-q^2-\frac{\epsilon^2}{\tilde{c}_{L(T)}^2}q^2\Delta \tilde{\Sigma}_{L(T)}^1(\textbf{q},\omega)|}$ (black line) and $\frac{1}{|\tilde{q}_{0L(T)}^2-q^2|}$ (dashed line) as a function of $aq$ for a given value of $aq_{0L(T)}$. The covariance function is cast in an exponential decay function. The values of the theory's input parameters are listed in the text.} \label{maxDeltaSig1}
\end{center}
\end{figure*}
We show that in the case of an exponential decay of the covariance function , as required by the hypotheses of Theorem I, for $q, \omega \in \mathbb{R}$,  i) $|\Delta \tilde{\Sigma}_{i i}^1(\textbf{q},\omega)|$ is continuous and bounded for $q \in [0,q_{Max}^{i}]$, $\exists q_{Max}^{i}$ $\forall i$; ii) $\mathrm{Im} [\Delta \tilde{\Sigma}_{i i}^1(\textbf{q},\omega)]>0$.
The self-energy in the Born Approximation, 
\begin{multline}
\tilde{\Sigma}_{i i}^1(\textbf{q},\omega)=\int_{-1}^{1} dx L_{i i j j}(x)\frac{2\pi }{c_{j}^2} \cdot \\ \cdot lim_{\eta \rightarrow 0^+}\int_{0}^{\infty} dq' q'^2 c(q,q',x) \frac{1}{q_{0 j,\eta}^2-q'^2},
\label{Sigma1_gen}
\end{multline}
can in such a case be calculated by exploiting the Sokhotski-Plemelj theorem and the Cauchy's Residue Theorem \cite{Turner,Calvet}, finding
\begin{widetext}
\begin{multline}
\tilde{\Sigma}_{i i}^1(\textbf{q},\omega)= i\int_{-1}^{1} dx L_{i i jj}(x) \frac{2}{c_{j}^2} \frac{(aq_{0 j})^3}{(1+(aq)^2+(aq_{0 j})^2+2(aq)(aq_{0 j})x)^2}- \int_{-1}^{1} dx L_{i i j j}(x) \frac{2}{c_{j}^2}  \frac{1}{\tilde{a}^3} \{[\tilde{a}^{10}+2(aq)^4x^4 \cdot \\ \cdot (-(aq_{0 j})^2+(aq)^2x^2)^3+ \tilde{a}^8(5(aq_{0 j})^2+ 6(aq)^2x^2)+a^6(7(aq_{0 j})^4+17(aq)^2(aq_{0 j})^2x^2+ 4(aq)^4x^4)+\tilde{a}^4(3(aq_{0 j})^6+ 16(aq)^2(aq_{0 j})^4x^2+\\13(aq)^4(aq_{0 j})^2x^4+16(aq)^6x^6)+\tilde{a}^2(-3(aq)^2(aq_{0 j})^6x^2 -(aq)^4(aq_{0 j})^4x^4-5(aq)^6(aq_{0 j})^2x^6+ 9(aq)^8x^8)]/[\tilde{a}^4+((aq_{0 j})^2-(aq)^2x^2)^2+\\2\tilde{a}^2((aq_{0 j})^2+(aq)^2x^2)^2]^2 \},
\label{SigmaLL_explicito_l1}
\end{multline}
\end{widetext}
where $\tilde{a}(q,x)=[(1-x^2)(aq)^2+1]^{1/2}$ and $L_{LL}(x)=4x^4$, $L_{LT}(x)=4(1-x^2)x^2$, $L_{TT}(x)=\frac{1}{2}(1-3x^2+4x^4)$, $L_{TL}(x)=2(1-x^2)x^2$.
The x-integration is performed numerically. Furthermore, it is $\tilde{\Sigma}_{i i}^1(0,0)=-\int_{-1}^{1} dx \frac{2}{c_{j}^2}L_{i i j j}(x) $. From inspection of Eq. \ref{SigmaLL_explicito_l1} we deduce that $\mathrm{Im}[ \Delta \tilde{\Sigma}_{i i}^1(\textbf{q},\omega) ]>0$; $|\Delta \tilde{\Sigma}_{i i}^1(\textbf{q},\omega)| \in C^0$ for $q,\omega \in \mathbb{R}$.

We define in the following the domain of the $(\omega, q)$ plane where Corollary II holds.
Specifics of the mathematical passages are outlined in Supplementary Note 1.
We first specify the domain of {convergence a.e. in} the $(\omega,q)$ plane of the series representation of $<G_{ii}(\textbf{q},\omega)>^1$. We then show that the magnitude of the remainder function, $|\tilde{R}_{k}(\textbf{q},\omega,\epsilon^2)|$, is as small as required by Corollary II when $aq_{Max}^{i}$ is large enough. We finally note that such a condition corresponds to deal with small values of $\frac{\epsilon^2}{\tilde{c}_{i}^2}$. 

\begin{figure}[h!]
\begin{center}
\includegraphics[scale=0.27]{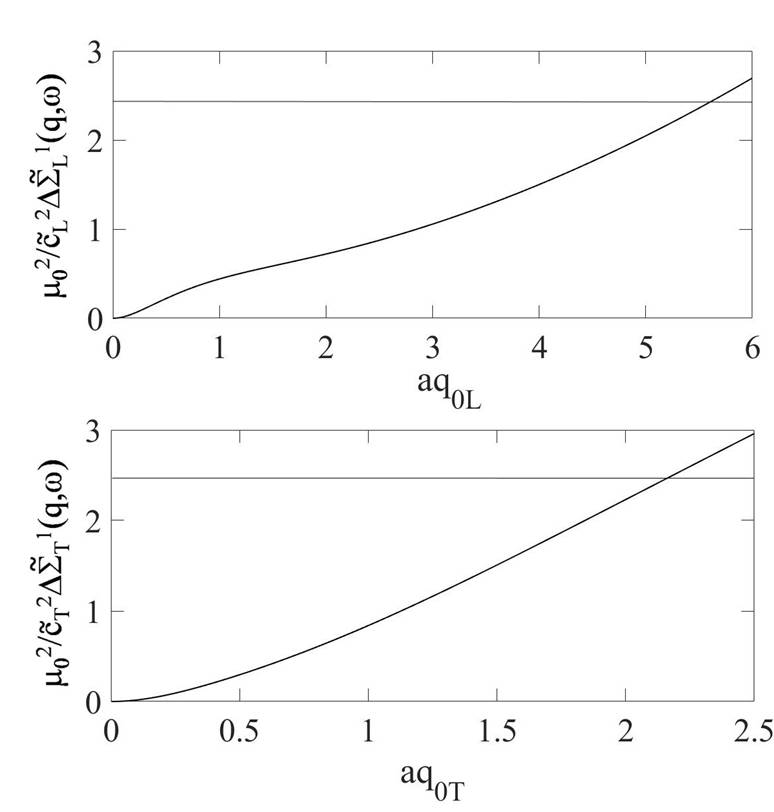}
\caption{$\frac{\mu_0^2}{\tilde{c}_{L(T)}^2}|\Delta \tilde{\Sigma}_{L}^1(q_{0L(T)},\omega)|$ for an exponentially decaying covariance function. The straight line fixes the value of $(\tilde{\epsilon}^2)^{-1}$. The values of the theory's input parameters are listed in the text.} \label{DeltaSigq0_q0}
\end{center}
\end{figure}

\begin{figure}[h!]
\begin{center}
\includegraphics[scale=0.3]{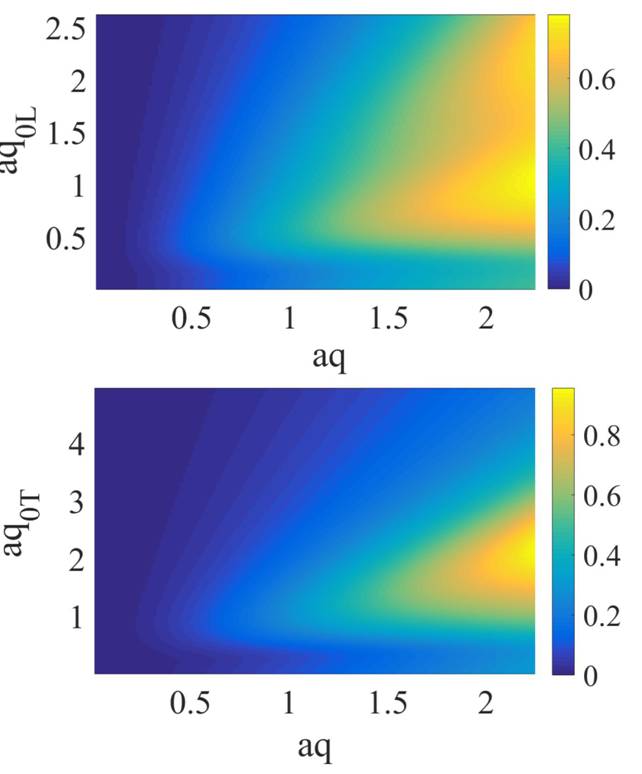}
\caption{Projection on the $(\omega,q)$ plane of $\frac{|\Delta \tilde{\Sigma}_{L}^1(q,\omega)-\Delta \tilde{\Sigma}_L^1(0,\omega)|}{|\Delta \tilde{\Sigma}_L^1(0,\omega)|}$ (\textit{Left Panel}) and $\frac{|\Delta \tilde{\Sigma}_{T}^1(q,\omega)-\Delta \tilde{\Sigma}_T^1(0,\omega)|}{|\Delta \tilde{\Sigma}_T^1(0,\omega)|}$ (\textit{Right Panel}) for an exponential decay of the covariance function. The values of the theory's input parameters are listed in the text.} \label{DeltaSig0}
\end{center}
\end{figure}

\begin{figure*}[h!]
\begin{center}
\includegraphics[scale=0.28]{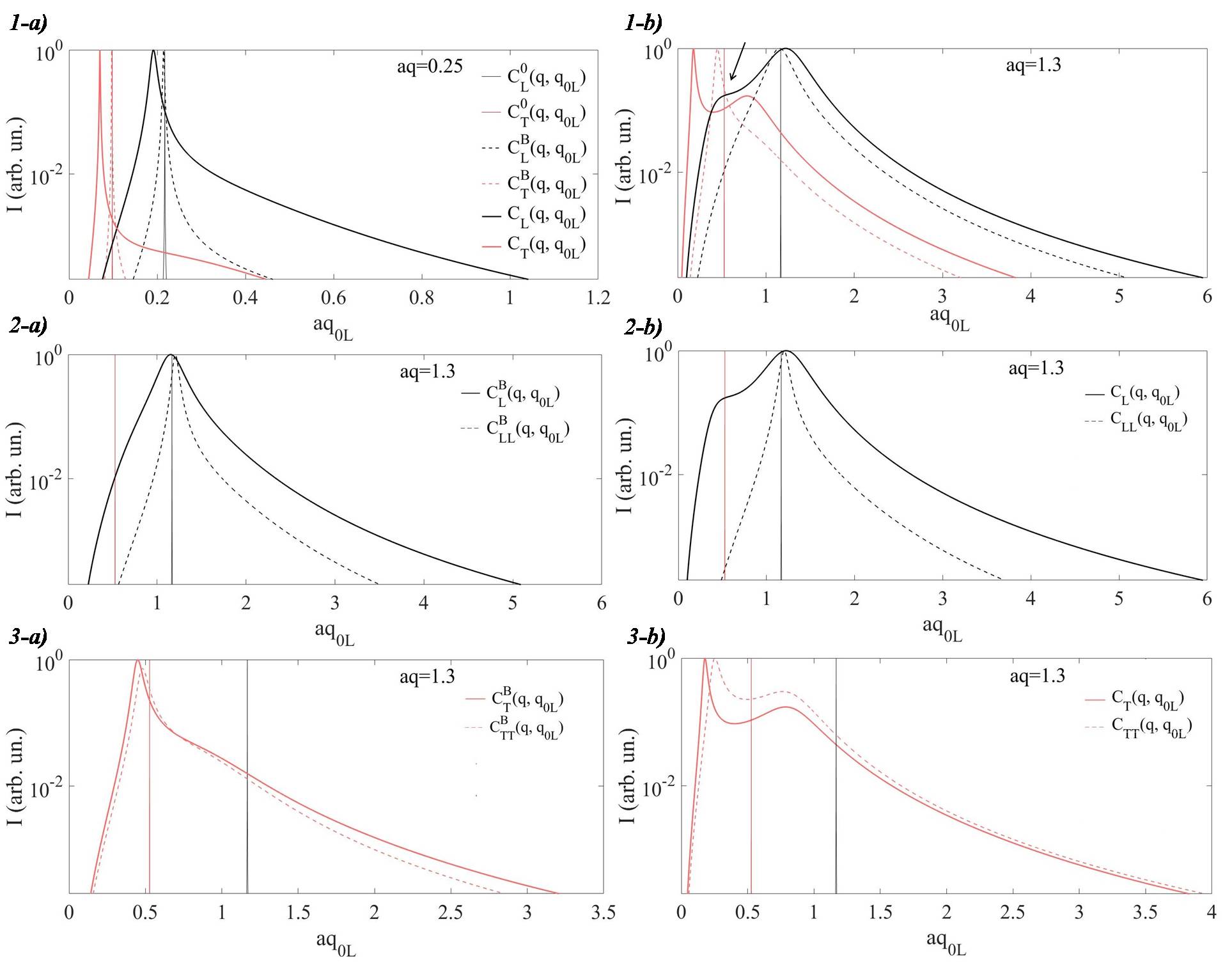}
\caption{\textit{Panels} 1. Longitudinal and transverse currents (black and red lines respectively) obtained by exploiting the GBA (full line) and the Born Approximation (dot-dashed line) in the case of an exponential decay of the covariance function for different values of wavevector. The values of the theory's input parameters, listed in the text, are the same for both approximations. \textit{Panels} 2. Longitudinal currents obtained by exploiting the Born Approximation, $a)$, and the GBA, $b)$. Full lines show the currents obtained by considering the full expression of the self-energy, $\Sigma_L(q,\omega)$, dot-dashed lines show the currents obtained by considering the only longitudinal contribution to the self-energy, $\Sigma_{LL}(q,\omega)$. \textit{Panels} 3. Transverse currents obtained by exploiting the Born Approximation, $a)$, and the GBA, $b)$. Full lines show the currents obtained by considering the full expression of the self-energy, $\Sigma_T(q,\omega)$, dot-dashed lines show the currents obtained by considering only the transverse contribution to the self-energy, $\Sigma_{TT}(q,\omega)$.} \label{Fig_correnti}
\end{center}
\end{figure*}

{As we} infer from Eq. \ref{SigmaLL_explicito_l1} and observe in Fig. \ref{maxDeltaSig1}, $|\Delta \tilde{\Sigma}_{i i}^1(\textbf{q},\omega)|$: i) definitively and independently from $\omega$ increases by increasing $q$ with a $q^2$ leading term ; ii) it has a local maximum in $q \sim q_{0i}$. The value of this maximum increases by increasing $q_{0i}$, as it is possible to observe in Fig. \ref{maxDeltaSig1}, {Panels} 2.
{It follows from} Theorem I {that} the condition $\frac{\epsilon^2}{\tilde{c}_{i}^2}\Delta \tilde{\Sigma}^{1,Max}_{i}(\omega)<1$  permits to discriminate the values of $q$ and $\omega$ where the series representation of $<G_{ii}(\textbf{q},\omega)>^1$ is a.e. convergent. {Given the properties of $|\Delta \tilde{\Sigma}_{i i}^1(\textbf{q},\omega)|$ stated in} points $i)$ and $ii)$ above we infer that this inequality is fulfilled {for values of $\omega$ and $q \in [0,q^i_{Max}]$:} a) $\frac{\epsilon^2}{\tilde{c}_{i}^2}|\Delta \tilde{\Sigma}^1_{i i}(q_{0i},\omega)|<1$ and, b) $\frac{\epsilon^2}{\tilde{c}_{i}^2}|\Delta \tilde{\Sigma}_{i i}^1(q_{Max}^i,\omega)|<1$. 
We observe that $|\Delta \tilde{\Sigma}^1_{i i}(q_{0i},\omega)|$ increases by increasing $q_{0i}$ with a $q_{0i}^3$ leading term, see Eq. \ref{SigmaLL_explicito_l1}. The frequency values where the condition a) is satisfied are thus $q_{0i} = \frac{\omega}{c^0_i}\ll q_{Max}^{i}$, as it is possible to infer also by the observation of Fig. \ref{DeltaSigq0_q0}.
We then fix a value of frequency where the conditions a) and b) are satisfied and observe that for $q\gg q_{0 i}$, i.e. $q \sim q_{Max}^{i} $, $\mathrm{Im}[ \Delta \tilde{\Sigma}_{i i}^1(\textbf{q},\omega) ] \ll 1$. This is verified in Supplementary Note 1. Consequently, the function $\frac{1}{|\tilde{q}_{0i}^2-q^2-\frac{\epsilon^2}{\tilde{c}_{i}^2}q^2\Delta \tilde{\Sigma}_{i i}^1(\textbf{q},\omega)|}$ has a local maximum at $\overline{q}^{i}:\frac{\epsilon^2}{\tilde{c}_{i}^2}|Re\{ \Delta \tilde{\Sigma}_{i i}^1(\overline{\textbf{q}}^{i},\omega)\}|=\frac{{\overline{q}^{i}}^2-\tilde{q}_{0i}^2}{{\overline{q}^{i}}^2}$. At larger wavevectors this function monotonically decreases by increasing $q$ until to be lower of $\frac{1}{|\tilde{q}_{0i}^2-q^2|}$. This behavior can be observed in Fig. \ref{DeltaSigq0_q0}, Panels 3.
The trends  outlined permit finally to asses that for $q_{0i}\ll q_{Max}^{i}$ and $q\ll Min_{\{ i \}}[q_{Max}^{i}]$ {and for values sufficiently large of $aq^i_{Max}$}, it is $|\tilde{R}_{k}(\textbf{q},\omega,\epsilon^2)| \lesssim \sum_{i} \frac{1}{\tilde{c}_{i}^2} \frac{1}{aq_{Max}^{i}}$. The mathematical passages are shown in detail in Supplementary Note 1. It is thus $|\tilde{R}_{k}(\textbf{q},\omega,\epsilon^2)|\ll 1$ when $aq_{Max}^{i}$ is large enough. In this case Corollary II holds and we can exploit the GBA. We infer from point i) and observe in Fig. \ref{maxDeltaSig1}, {Panels} 2, that as smaller it is $\frac{\epsilon^2}{\tilde{c}_{i}^2}$ as larger is $aq_{Max}^{i}$. It furthermore follows that for a given value of $\epsilon^2$ the larger it is $\tilde{c}_i$, the larger $aq^i_{Max}$. The magnitude of the remainder function remain thus uniquely linked to the value of $\frac{\epsilon^2}{\tilde{c}_{i}^2}$. Finally, noting the inverse proportionality between $aq_{Max}^i$ and the magnitude of the remainder function, we expect that the domain of the $(\omega,q)$ plane where the GBA can be applied includes values of $q:aq \sim 1$.  A specific case is treated in Figs. \ref{DeltaSigq0_q0} - \ref{Fig_correnti}, where the quantities shown have been obtained for theory's input parameters which allow to describe the features of longitudinal dynamics for a real system \cite{izzo}. They are $c^0_L=2.29$ $meV/nm^{-1}$, $c^0_T/c^0_L=0.53$, $\tilde{\epsilon}^2=\frac{\epsilon^2}{\mu_0^2}=0.4$ and $a=1.1$ $nm$. By inspection of Figs. \ref{DeltaSigq0_q0}-\ref{DeltaSig0} we find $aq_{Max}^L=18$, $aq_{Max}^T=8$. For such theory's input parameters the GBA hence holds for $q:aq \ll 8$. 
The mixing of polarization is expected for $aq \sim 1$. In this case the proposed approximation can thus be used to describe such a phenomenon. 

For the input parameters specified above we calculate the longitudinal and transverse self-energies in the GBA. In Sec. \ref{mixing} and \ref{acoustic_prop} we analyse the features {of the acoustic dynamics} focusing, in particular, on the mixing of polarizations. 
To this aim we truncate the series in Eq. \ref{serieCo} to the order $n=1$, obtaining $\Sigma_{kk}(\textbf{q},\omega) \sim F_k^0(\textbf{q},\omega)+F_k^1(\textbf{q},\omega)$. In Supplementary Note 2 we numerically retrieve the value of $|F_k^1(\textbf{q},\omega)|$ and $|\tilde{R}_k(\textbf{q},\omega,\epsilon^2)|$ for selected values of wavevector and frequency and show that the former is significantly larger than the latter. We notice that the series in Eq. \ref{serieCo} is obtained as the integral of a power series a.e. convergent. It is thus expected that the leading-order terms will be the ones with smaller $n$.
Such an order of approximation is furthermore sufficient to obtain a realistic description of the acoustic dynamics, including the Rayleigh anomalies and the mixing of polarizations \cite{izzo}. In order to facilitate such calculation we furthermore assume $\Delta \tilde{\Sigma}_{i i}^1(\textbf{q}',\omega) \sim \Delta \tilde{\Sigma}_{i i}^1(0,\omega)$.
When the approximation $\Delta \tilde{\Sigma}_{i i}^1(\textbf{q}',\omega) \sim \Delta \tilde{\Sigma}_{i i}^1(0,\omega)$ applies, the Hadamard principal value of the integral defining $F_k^1(\textbf{q},\omega)$ can be obtained straightforwardly by exploiting the Residue Theorem.
In such a case we can indeed extend the upper integration boundary of the integral to infinity while maintaining unaffected the order of magnitude of the error related to the GBA, as discussed in Supplementary Note 3. 
It is furthermore shortly discussed in Supplementary Note 4 that as long as the condition $|\frac{\Delta \tilde{\Sigma}_{i i}^1(\textbf{q}',\omega)-\Delta \tilde{\Sigma}_{i i}^1(0,\omega)}{\Delta \tilde{\Sigma}_{i i}^1(0,\omega)}|<\frac{1}{2}$ is fulfilled the dominant contribution to the integral defing $F_k^1$ can be obtained trough the approximation $\Delta \tilde{\Sigma}_{i i}^1(\textbf{q}',\omega) \sim \Delta \tilde{\Sigma}_{i i}^1(0,\omega)$.  Fig. \ref{DeltaSig0} shows $|\frac{\Delta \tilde{\Sigma}_{i i}^1(\textbf{q}',\omega)-\Delta \tilde{\Sigma}_{i i}^1(0,\omega)]}{\Delta \tilde{\Sigma}_{i i}^1(0,\omega)}|$ for an exponential decay of the covariance function and for the given input parameters. This condition is fulfilled up to frequencies and wavevectors $aq^{L(T)}(q_{0L(T)}) \sim 2$.  Since the shape of the covariance function makes that the larger contribution to the integral defining $F_k^1$ is for $q' \sim q \pm a^{-1}$, the approximation $\Delta \tilde{\Sigma}_{i i}^1(\textbf{q}',\omega) \sim \Delta \tilde{\Sigma}_{i i}^1(0,\omega)$ is assumed to give a significant estimation of the integral up to wavevectors of the order of $a^{-1}$, where we aim to focus in the present study.
\begin{figure*}[h!]
\begin{center}
\includegraphics[scale=0.45]{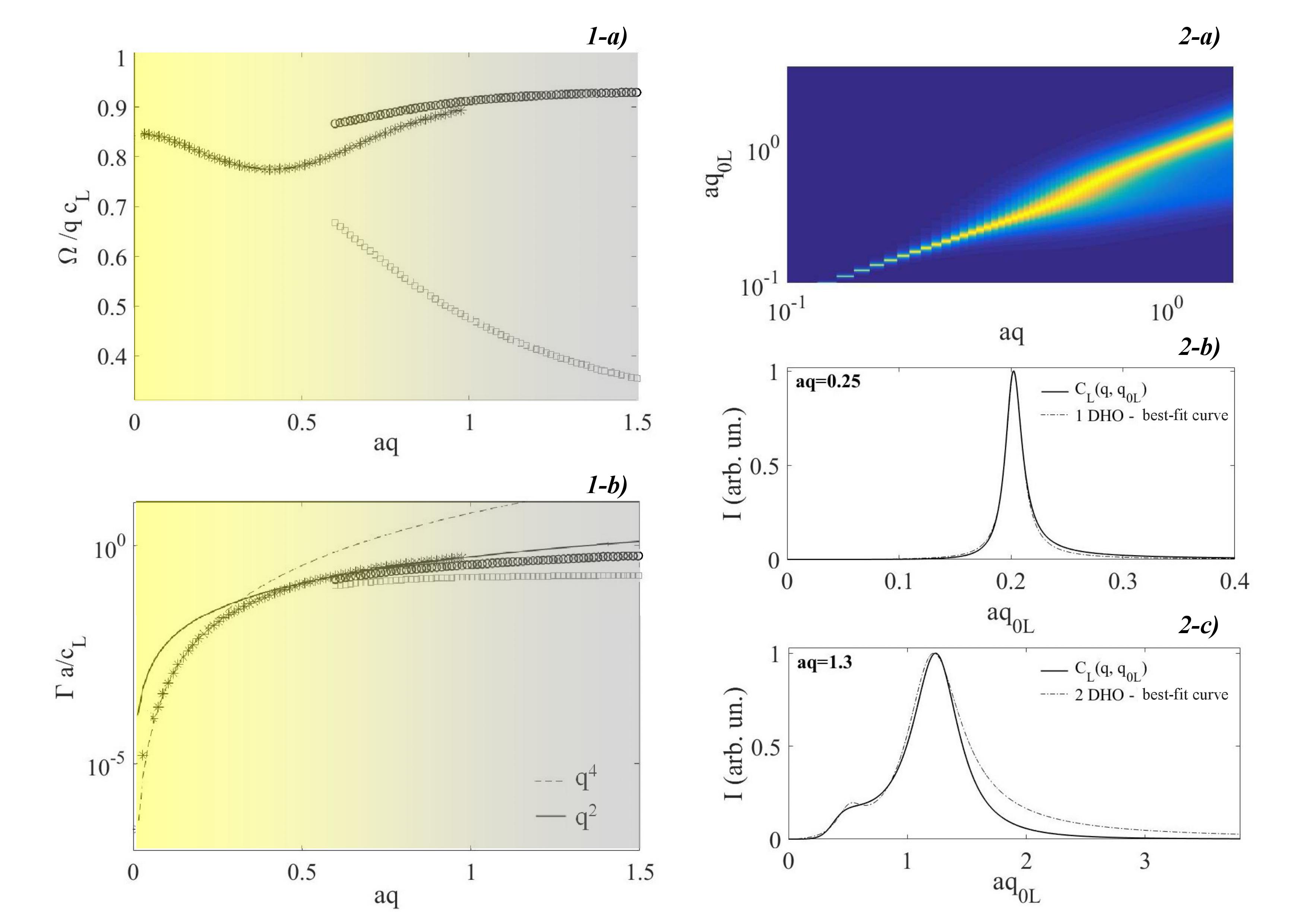}
\caption{{Features of the longitudinal acoustic dynamics obtained by exploiting the GBA both in the Rayleigh region, $aq<<1$ and in the wavevectors region $aq \sim 1$ where the mixing of polarizations shows up. \textit{Panel 1-a)}. Adimensional phase velocity $\Omega / qc_L$ as a function of $aq$. Stars represent outcomes from 1-DHO model fitting, circles and squares  represent outcomes from 2-DHO model fitting related respectively to high- and low-frequency features. \textit{Panel 1-b)}. Adimensional attenuation $\Gamma a / c_L$ as function of $aq$. The meaning of the symbols is the same of Panel 1-a). Full black line and dashed line are guide to eye displaying respectively the $q^4$ and $q^2$ trend. \textit{Panel 2-a)}. Projection on the $(aq, aq_{0L})$ plane of the longitudinal currents. \textit{Panel 2-b)} Representative longitudinal current spectrum in the Rayleigh region (full black line). The dashed line shows the best-fit curve with a 1-DHO fitting model. \textit{Panel (2-c))} Representative longitudinal current spectrum in the wavevector region where the mixing of polarizations is clearly observable (full black line). The dashed line shows the best-fit curve with a 2-DHO fitting model.}} \label{Rmix}
\end{center}
\end{figure*}
Fig. \ref{Fig_correnti}, {Panels} $1$, shows the longitudinal and transverse currents (black and red bold lines respectively) obtained by exploiting the GBA for two different values of wavevector $a)$, $aq\ll 1$ and, $b)$, $aq \sim 1$. The maximum of the current is normalized to one. The current $C_{L(T)}(q,\omega)$ is obtained from the dynamic structure factor $S_{L(T)}(q,\omega)$, being $C_{L(T)}(q, \omega)=\frac{\omega^2}{q^2}S_{L(T)}(q, \omega)$. The dynamic structure factors are related to the average Green functions through the fluctuation-dissipation theorem, $S_{L(T)}(q, \omega) \propto \frac{q^2}{\omega} \mathrm{Im} [<g_{L(T)}(q,\omega)>]$. 
The longitudinal and transverse self-energies defining the average Green functions are obtained from the GBA as described in Sec. \ref{domain}. The longitudinal and transverse currents calculated by exploiting the Born Approximation (dot-dashed lines) are furthermore shown together with the currents of a \lq bare' medium with phase velocities equal to the average first-order perturbed phase velocities, $\tilde{c}_{L(T)}$. They are shown respectively as black and red straight lines and referred to be $C^0_{L(T)}(q,\omega)$.
\subsection{The mixing of polarizations in the Born and Generalized Born Approximations} \label{mixing}
The main peak in the longitudinal currents obtained from the GBA points the inelastic excitation centered at the characteristic frequency determined by the longitudinal phase velocity. For the only case $aq \sim 1$ it is furthermore {observed} a low-frequency feature, which can be described as a secondary peak centered at frequencies characteristic of the transverse excitations. This kind of feature has been observed experimentally and by Molecular Dynamics (MD) simulations in several topologically disordered systems \cite{Ruzi, Scopigno1HFglasses_transverse, zanatta_INS_GeO2,Cunsolo, Bolmotov1, Sampoli, Bryk1,Ribeiro2,Bolmotov2}. In the current spectrum related to the Born Approximation we observe a shoulder-like feature at the same frequency, but such a feature is clearly more pronounced when the GBA is used. 
The endorsement of the fact that the secondary peak is related to the mixing of polarizations comes from the fact that it disappears when the cross term accounting for the coupling with transverse dynamics, $\Sigma_{LT}(q,\omega)$, is removed from the longitudinal self-energy. {This is emphasized} in Fig. \ref{Fig_correnti}, {Panels} $2$ , where they have been shown the currents obtained by using {respectively} the full expression of the self-energy, $\Sigma_L(q,\omega)$, {(full line)} and the only longitudinal contribution to the self-energy, $\Sigma_{LL}(q,\omega)$, {(dot-dashed line)}. Both in the case of the Born Approximation and of the GBA the features observed at the characteristic frequencies of the transverse excitations, which are present when the full expression of the self-energy is taken under account, disappear when the only term related to the longitudinal contribution to the self-energy is considered. This is not the case when we are dealing with the transverse dynamics, as it is possible to infer by observing Fig. \ref{Fig_correnti}, {Panels} $3$. In this case the secondary peak observed at frequencies higher than the one defined by the transverse phase velocity is in part preserved when the longitudinal contribution to the self-energy is left out. The occurrence of the secondary peak can be related to the existence of a two-modes regime, observed in a random media whose covariance function can be described as $i)$ an exponential decay function \cite{Calvet} for wavevectors and frequencies $aq(q_{0 i})>1$. This behavior can be reproduced also by using the scalar Born Approximation \cite{B1,Calvet}. Similar feature is furthermore observable in the longitudinal dynamics for wavevectors higher than the ones considered in Fig. \ref{Fig_correnti}. It can coexist with the feature related to the mixing of polarization. 
\subsection{Features of the acoustic dynamics in the Generalized Born Approximation} \label{acoustic_prop}
{The features of the longitudinal acoustic dynamics obtained by GBA up to wavevector of the order of $a^{-1}$ have been derived from the calculated longitudinal currents by a fitting procedure described in the following and qualitatively compared with experimental finding reported in the literature. The dynamic structure factors related to longitudinal acoustic dynamics in topologically disordered systems have been experimentally characterized by several studies mostly based on IXS or INS measurements in different wavevectors regions. An universal behavior emerged, which can be qualitatively described by i) the presence of the so-called Rayleigh anomalies for values of wavevctors lower than $a^{-1}$, i.e. the phase velocity of the acoustic excitations shows a softening with respect to its macroscopic value while the acoustic mode attenuation is affected by a strong increase and roughly follows a $q^4$ trend; ii) increase of the phase velocity at higher wavevector values, resulting in a minimum in its wavevector-trend, and crossover from $q^4$ to $q^2$ trend of the acoustic modes attenuation \cite{hydro_MonGio, Ruta, Baldi, Mossa, Ferrante}; iii) mixing of polarizations for $q \sim a^{-1}$ manifesting in the presence of a peak-like feature in the longitudinal current spectra at the characteristic frequencies of transverse excitations \cite{Ruzi, Scopigno1HFglasses_transverse, Cunsolo, Bolmotov1, zanatta_INS_GeO2, Bencivenga, Cimatoribus}. The GBA can grasp all these characteristics. While the Rayleigh anomalies can be obtained also by exploiting the Born Approximation \cite{John, Calvet} or the SCBA \cite{Schirmacher_4th,Marruzzo,Ferrante}, the mixing of polarizations can be accounted only by the GBA. It is worth, furthermore, to observe that at the boundary of the Rayleigh region when depolarization effects begin to affect the acoustic dynamics, the coupling between longitudinal and transverse dynamics, though not manifesting in a clear peak-like feature in the longitudinal currents, can have an impact on the effective experimentally observed attenuation and phase velocity. To obtain a realistic description of the acoustic dynamics also in this wavevectors region it is thus more appropriate to consider a vectorial model in the RMT frame, such as the GBA.   
The features of the longitudinal acoustic dynamics can be derived from the longitudinal currents obtained by the GBA by fitting the calculated spectra with a fitting model composed by one or two Damped Harmonic Oscillator (DHO) functions, following the same protocol usually used to analyze the experimental IXS or INS data. This approach, also referred to be as spectral function approach, has also been used in the analysis of theoretical results aimed to characterize the acoustic dynamics in random media \cite{Calvet, Sheng}. Because most of the experimental data presented in literature have been analyzed with the above quoted fitting model, the spectral function approach permits a clear connection between the GBA theoretical outputs and the literature results. 
The longitudinal currents produced by the GBA have been modeled with the expression
\begin{eqnarray}
& C_L(q,\omega)=\omega^2\sum_{n} \frac{I_{_{n}}\Gamma_{n} \Omega_{n}^2}{(\Omega_{n}^2-\omega^2)^2+\omega^2 \Gamma_{n}^2} \ \
\end{eqnarray}
where $n=1$ in the Rayleigh region $q<<a^{-1}$ and $n=1,2$ in the region $q \sim a^{-1}$. In the transition region, where the feature related to the tranverse dynamics starts to show up, the fitting has been performed by exploiting both 1- or 2-DHO model fit functions. The phase velocity, $c(q)$ is related to the characteristic frequency, $\Omega$, trough the relationships $c(q)=\frac{\Omega(q)}{q}$, while the parameter $\Gamma$ is directly related to the acoustic mode attenuation. The wavevector-dependent outcomes of the fitting procedure are diplayed in Fig. \ref{Rmix}. The theory's input parameters are the same listed above. Both the Rayleigh anomalies and the mixing of polarizations are clearly observed in qualitative agreement with most of experimental outcomes reported in literature. A quantitative comparison with experimental outcomes is reported in Ref. \cite{izzo}.}
 
\section{Conclusion} \label{conclusion}
By introducing corrective terms to the Born Approximation we obtained in an analytic form an expression for the self-energy related to the stochastic Helmholtz equation describing the acoustic dynamics in an elestically heterogeneous medium. {In the frame of} the perturbative series expansion of the Dyson equation the proposed approximation accounts in an approximate form up to the second order term, whereas the Born Approximation stops to the first order. The Feynman diagram technique permits to clarify which multiple scattering events are {included} in the Generalized Born Approximation. 
The case of a covariance function given by an exponential decay function is analysed in some detail. In such a case it was proved the validity of the proposed approximation in a domain of the $(\omega, q)$ plane of interest in most topologically disordered systems (e.g. glasses). It includes both the Rayleigh region and wavevectors {region: $aq \sim 1$}, where it is expected the mixing of polarizations to get in.
Furthermore, the validity of the Generalized Born Approximation is not restricted to $q$ in a neighbor of $\frac{\omega}{c_{i}}$, where $c_{i}$ is the phase velocity of the unperturbed medium for the $i$-th polarization, thus permitting to describe also features of the average Green's dyadic occurring at frequencies smaller or higher than $c_iq$, as it is the case for the mixing of polarizations. 
We finally verified that the proposed approximation permits to describe this phenomenon {together with the Rayleigh anomalies.}
% The error related to the proposed approximation is evaluated. 

Acoustic modes with mixed polarization have been observed by both IXS and INS as well as by MD simulations in several disordered systems. The phenomenon has never been related, however, to Rayleigh anomalies and quantitatively described as a phenomenon also originating from the disordered nature of the medium. The proposed approximation can permit to reach this goal and to trace the way towards a coherent and experimentally verifiable mathematical description of all the phenomena arising from the elastic heterogeneous structure of an amorphous solid.

\section*{Conflict of Interest Statement}
%All financial, commercial or other relationships that might be perceived by the academic community as representing a potential conflict of interest must be disclosed. If no such relationship exists, authors will be asked to confirm the following statement: 

The authors declare that the research was conducted in the absence of any commercial or financial relationships that could be construed as a potential conflict of interest.

\section*{Author Contributions}

M. G. I. led research, produced the results and wrote the paper.
G. R. discussed the results and revisited the paper.
S. C. discussed the results and revisited the paper.

%\section*{Funding}
%Details of all funding sources should be provided, including grant numbers if applicable.
% Please ensure to add all necessary funding information, as after publication this is no longer possible.

\section*{Acknowledgments}
The authors acknoweldge W. Schirmacher and G. Pastore for usefull discussions.

%
% \pnasbreak splits and balances the columns before the references.
% If you see unexpected formatting errors, try commenting out this line
% as it can run into problems with floats and footnotes on the final page.
%\pnasbreak
%
%

% Bibliography
\bibliography{test}

\clearpage
\onecolumn
\section{\textbf{Appendix A}} \label{appendixA}
\renewcommand\theequation{A\arabic{equation}}
\setcounter{equation}{0}
We show that 
\begin{eqnarray}
\frac{1}{N!}\frac{d^N}{dz^N}\frac{[z^2]^N}{(z+\tilde{q}_{0i})^{N+1}}|_{z=\tilde{q}_{0i}} =\frac{N+1}{2N+1} \frac{1}{\tilde{q}_{0 i}}.
\end{eqnarray}
For a generic product of functions $f(z)g(z)$, it is
\begin{eqnarray}
\frac{d^N}{dz^N} [f(z)g(z)]=\sum_{k=0}^N \frac{N!}{k!(N-k)!}\frac{d^k}{dz^k}f(q)\frac{d^{N-k}}{dz^{N-k}}g(z).
\label{derivatan_prodottofunzioni_1}
\end{eqnarray} 
Furthermore
\begin{itemize}
\item $\frac{d^k}{dz^k} z^{2N}=\frac{(2N)!}{(2N-k)!} z^{2N-k}$; 
\item $\frac{d^{N-k}}{dz^{N-k}}(z+\tilde{q}_{0 i})^{-(N+1)}=(-1)^{N-k}\frac{(2N-k)!}{N!} (z+\tilde{q}_{0 i})^{-(2N-k+1)}$,
\end{itemize}
from which we obtain
%\begin{widetext}
\begin{multline}
[ \frac{d^k}{dz^k} z^{2N}\frac{d^{N-k}}{dz^{N-k}}(z+\tilde{q}_{0 i})^{-(N+1)}]|_{z=\tilde{q}_{0 i}} =  (-1)^{N-k}\frac{(2N)!}{N!}  2^{-(2N-k+1)} \tilde{q}_{0 i}^{-1}=\\ (-1)^{N-k} \frac{(2N+1)!(N+1)}{2^{N+1}(N+1)!(2N+1)} 2^{-(N-k)} \tilde{q}_{0 i}^{-1}= (-1)^{N-k}\frac{(2N)!!(N+1)}{(2N+1)} 2^{-(N-k)} \tilde{q}_{0 i}^{-1} = (-1)^{N-k}\frac{2^NN!(N+1)}{(2N+1)} 2^{-(N-k)}\tilde{q}_{0 i}^{-1},
\label{derivatan_prodottofunzioni_2}
\end{multline} 
%\end{widetext}
where the relationships $2^N N!=(2N)!!$ has been exploited, being $n!!$ the double factorial function of the integer n. It follows
\begin{multline}
\frac{1}{N!}\sum_{k=0}^N \frac{N!}{k!(N-k)!} [ \frac{d^k}{dz^k} z^{2N}\frac{d^{N-k}}{dz^{N-k}}(z+\tilde{q}_{0 i})^{-(N+1)}]|_{z=\tilde{q}_{0 i}} =\sum_{k=0}^N \frac{2^N(N+1)}{(2N+1)}\tilde{q}_{0 i}^{-1}\frac{N!}{k!(N-k)!}(-1)^{N-k}2^{(N-k)}= \frac{N+1}{2N+1} \tilde{q}_{0 i}^{-1},
\label{derivatan_prodottofunzioni_3}
\end{multline} 
being from the binomial theorem $ \sum_{k=0}^N\frac{N!}{k!(N-k)!}(-1)^{N-k}2^{-(N-k)}=\sum_{k=0}^N \frac{N!}{k!(N-k)!}(-\frac{1}{2})^{-(N-k)}=(1-\frac{1}{2})^N$. 

\section{\textbf{Appendix B}} \label{appendixB}
\renewcommand\theequation{B\arabic{equation}}
\setcounter{equation}{0}
{Truncation at the second order of the perturbative series expansion can give an approximate expression of the self-energy when the necessary condition $|\mathbf{\Sigma}^{3}(\textbf{q},\omega)-\mathbf{\Sigma}^{2}(\textbf{q},\omega)|\ll |\mathbf{\Sigma}^{2}(\textbf{q},\omega)|$ is satisfied.}  {Because under the hypothesis of local isotropy the self-energy dyadic is diagonal, this inequality is verified if $|\Sigma_{kk}^{3}(\textbf{q},\omega)-\Sigma_{kk}^{2}(\textbf{q},\omega)|\ll |\Sigma_{kk}^{2}(\textbf{q},\omega)|$.}
In the following we show that the latter inequality holds inside the domain of the $(\omega,q)$ plane where the series representation introduced in Sec. \ref{genborn} approximates the quantity $\mathbf{\Sigma}^2(\textbf{q},\omega)$ if the magnitude of the remainder function of order one of the series representation is small enough. We furthermore show that the necessary condition of validity for the GBA is less stringent than for the Born Approximation.

{It is $|\Sigma_{kk}^{3}(\textbf{q},\omega)-\Sigma_{kk}^{2}(\textbf{q},\omega)|=|\int_{-1}^1L_{kk i i}(x) \frac{1}{c_{i}^2} \int_0^{\infty} dq' \ q'^2 c(aq,aq',x) [\frac{1}{q_{0i}^2-q'^2-c_{i}^{-2}\Sigma_{i i}^{2}(\textbf{q}',\omega)}-\frac{1}{q_{0i}^2-q'^2-c_{i}^{-2}\Sigma_{i i}^{1}(\textbf{q}',\omega)}]|$.} Since as stated in Theorem I $\mathrm{Im} [ \tilde{\Sigma}_{i i}^{1}(\textbf{q},\omega)]>0$, the function $\frac{1}{q_{0i}^2-q'^2- c_{i}^{-2}\Sigma_{i i}^{2}(\textbf{q}',\omega)}$ can be represented as a Taylor series of argument $\Sigma_{i i}^{2}(\textbf{q},\omega)-\Sigma_{i i}^{1}(\textbf{q},\omega)$. The integral of the zero-th order term gives $\Sigma_{kk}^{2}(\textbf{q},\omega)$. {As smaller it is the argument of the Taylor series, as quickly the series converges and as smaller it is the remainder function with respect to the series truncation at a given order.} Under the hypotheses of validity of Theorem I and Corollaries I and II and by assuming that the intensity of spatial fluctuations are small enough so that $\tilde{q}_{0 i} \approx q_{0 i}$, the quantity $\Sigma_{i i}^{2}(\textbf{q},\omega)-\Sigma_{i i}^{1}(\textbf{q},\omega)$ can be approximated by the remainder function of order one of the series representation for $\Sigma_{ii}^{2}(\textbf{q},\omega)$ introduced in Sec. \ref{validity}, {which we call $S_{1}^{i}$.}
The condition $|\Sigma^{3}(\textbf{q},\omega)-\Sigma^{2}(\textbf{q},\omega)|\ll |\Sigma^{2}(\textbf{q},\omega)|$ is thus verified when $|S_{1}^{i}| \ll 1$.

A necessary condition for the validity of the Born Approximation is $|\Sigma_{kk}^{2}(\textbf{q},\omega)-\Sigma_{kk}^{1}(\textbf{q},\omega)|\ll |\Sigma_{kk}^{1}(\textbf{q},\omega)|$. Under the hypotheses of validity of Theorem I and Corollaries I and II and still by assuming $\tilde{q}_{0 i} \approx q_{0 i}$ this condition is equivalent to a quick convergence of the series development for $\mathbf{\Sigma}^2(\textbf{q},\omega)$ introduced in the text, i.e $|\tilde{S}_{1}^{i}(\textbf{q},\omega)| \ll |\tilde{\Sigma}^{n=1}(\textbf{q},\omega)|$. {It is defined $\tilde{S}_{1}^{i}(\textbf{q},\omega)=(\epsilon^2q^2)^{-1}S_{1}^{i}(\textbf{q},\omega)$, with}
\begin{multline}
\tilde{S}_{1}^{i}(\textbf{q},\omega)= lim_{\eta \rightarrow 0^+} \int_{0}^{q_{Max}^{i}}dq' \ q'^2 \ c(q,q',x) \frac{1}{(\tilde{q}_{0 i,\eta}^2-q'^2)} \frac{[\frac{\epsilon^2}{\tilde{c}_{i}^2}q'^2\Delta \tilde{\Sigma}_{ii}^{1}(\textbf{q}',\omega_{\eta})]^2}{(\tilde{q}_{0 i,\eta}^2-q'^2)[\tilde{q}_{0 i,\eta}^2-q'^2-\frac{\epsilon^2}{\tilde{c}_i^2}q'^2\Delta \tilde{\Sigma}_{ii}^{1}(\textbf{q}',\omega_{\eta})]}. 
\label{Rstar}
\end{multline}
We observe that
\begin{multline}
|\tilde{S}_{1}^{i}(\textbf{q},\omega)|\leq lim_{\eta \rightarrow 0^+}\int_{0}^{q_{Max}^{i}}dq' \ q'^2 \ c(q,q',x) \frac{1}{|\tilde{q}_{0 i,\eta}^2-q'^2|}\cdot  \tilde{M}^{i}\leq  2 |lim_{\eta \rightarrow 0^+}\int_{0}^{q_{Max}^{i}}dq' \ q'^2 \ c(q,q',x) \frac{1}{(\tilde{q}_{0 i,\eta}^2-q'^2)}|\cdot  \tilde{M}^{i} \sim \\ 2|\tilde{\Sigma}^{n=1}_{ii}(\textbf{q},\omega)| \tilde{M}^{i}, 
\label{Rstar_1}
\end{multline}
where $\tilde{M}^{i}=\sup_{q' \in [0,q_{Max}^i]} \big[\frac{[\frac{\epsilon^2}{\tilde{c}_{i}^2}q'^2\Delta \tilde{\Sigma}_{ii}^{1}(\textbf{q}',\omega_{\eta})]^2}{(\tilde{q}_{0 i,\eta}^2-q'^2)[\tilde{q}_{0 i,\eta}^2-q'^2-\frac{\epsilon^2}{\tilde{c}_i^2}q'^2\Delta \tilde{\Sigma}_{ii}^{1}(\textbf{q}',\omega_{\eta})]}\big]$. The second inequality in Eq. \ref{Rstar_1} is obtained by similar passages leading to Eqs. \ref{int_ex_3_0} and \ref{int_ex_3_1} with $N=0$ being $c(q,q',x) \in \mathbb{R}^+$ and considering that $|\mathrm{Re}[z]|+|\mathrm{Im}[z]|\leq 2 |z|$, where z is a complex number. In the case of the Born Approximation it is thus required that $\tilde{M}^{i} \ll 1$. In the case of the GBA the necessary condition of validity requires $|\tilde{\Sigma}^{1}_{ii}(\textbf{q},\omega)| \tilde{M}^{i} \ll 1$. This condition is less stringent than the former because in the case of small fluctuations when $\Delta \tilde{\Sigma}_{ii}(\textbf{q},\omega) \sim \tilde{\Sigma}_{ii}(\textbf{q}, \omega)$ it is $|\tilde{\Sigma}^{n=1}_{ii}(\textbf{q},\omega)| < 1$ under the conditions of validity of Theorem I.

\clearpage
\onecolumn
\section{Supplementary Note 1}
\renewcommand\theequation{S\arabic{equation}}
\setcounter{equation}{0}
We show that in the case of an exponential decay of the covariance function it is  $|\tilde{R}_{k}(\textbf{q},\omega,\epsilon^2)| \lesssim \sum_{i} \frac{1}{\tilde{c}_{i}^2} \frac{1}{aq_{Max}^{i}}$ in the domain of the $(\omega,q)$ plane specified in Sec. 3.2 of the main text , i.e. $q_{0i}\ll  q_{Max}^{i}$ and $q\ll Min_{\{ i \}}[q_{Max}^{i}$. 

\begin{figure}[tbp]
\begin{center}
\includegraphics[scale=0.18]{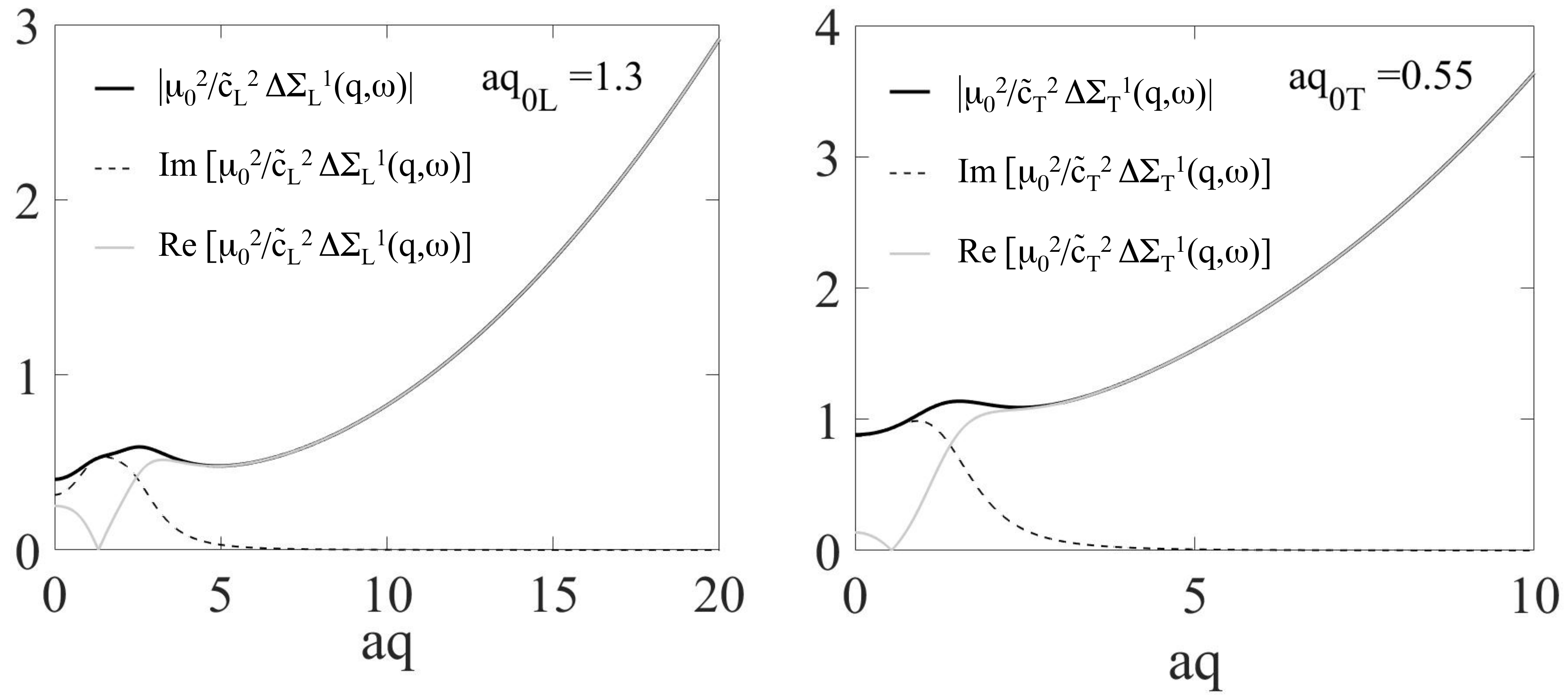}
\caption{$\frac{\mu_0^2}{\tilde{c}_{L(T)}^2}|\Delta \tilde{\Sigma}_{L(T)}^1(\textbf{q},\omega)|$ (black line), $\mathrm{Im}[\frac{\mu_0^2}{\tilde{c}_{L(T)}^2} \Delta \tilde{\Sigma}_{L(T)}^1(\textbf{q},\omega)]$ (dashed line) and $|\mathrm{Re}[ \frac{\mu_0^2}{\tilde{c}_{L(T)}^2}\Delta \tilde{\Sigma}_{L(T)}^1(\textbf{q},\omega)]|$ (grey line) as a function of $q$ for a fixed value of $aq_{0L(T)}\ll  aq_{Max}^{L(T)}$. The covariance function is an exponential decay function. The theory's input parameters are listed in the main text.} \label{absDeltaSig1_altiq}
\end{center}
\end{figure}

It can be inferred from Eq. 37 in the main text that for $q\gg q_{0 i}$ (i.e. $q \sim q_{Max}^{i}$),
\begin{enumerate}
\item [\textit{a})] $\tilde{q}_{0i}^2-q^2<0$ ;
\item [ \textit{b})] $Re \{\Delta \tilde{\Sigma}_{i}^1(\textbf{q},\omega)\}<0$; 
\item [\textit{c})] $|Re \{\Delta \tilde{\Sigma}_{i}^1(\textbf{q},\omega)\}|$ definitively increases by increasing $q$ with  a $q^2$ leading term (see Fig. 2);
\item [\textit{d})] $Im \{\Delta \tilde{\Sigma}_{i}^1(\textbf{q},\omega)\}>0$;
\item [\textit{e})] $Im \{\Delta \tilde{\Sigma}_{i}^1(\textbf{q},\omega)\}\ll  1$.
\end{enumerate}
Point \textit{a}) is valid because $q_{0i} \sim \tilde{q}_{0i}$ when $\epsilon^2$ is small. \textbf{To} support of point \textit{e}), Fig. \ref{absDeltaSig1_altiq} shows the wavevector trend of $|Re\{ \Delta \tilde{\Sigma}_{i}^1(\textbf{q},\omega)\}|$, $|Im\{ \Delta \tilde{\Sigma}_{i}^1(\textbf{q},\omega)\}|$ and $|\Delta \tilde{\Sigma}_{i}^1(\textbf{q},\omega)|$ for a given value of $q_{0i}\ll  q_{Max}^{i}$. For $q \sim q_{Max}^{i}$ we observe that the value of $|Im\{ \Delta \tilde{\Sigma}_{i}^1(\textbf{q},\omega)\}|$ is negligible with respect to the value of $|Re\{ \Delta \tilde{\Sigma}_{i}^1(\textbf{q},\omega)\}|$. 
We will call the wavevectors region \textbf{$:q \sim q_{Max}^{i}$}, the \lq high-$q$' region.

In the \lq high-$q$' region, 
\begin{enumerate}
\item [i)] the function $\frac{1}{|\tilde{q}_{0i}^2-q^2-\frac{\epsilon^2}{\tilde{c}_{i}^2}q^2\Delta \tilde{\Sigma}_{i}^1(\textbf{q},\omega)|}$ has a maximum in $\overline{q}_{i}:\frac{\epsilon^2}{\tilde{c}_{i}^2}|Re\{ \Delta \tilde{\Sigma}_{i}^1(\overline{\textbf{q}}_{i},\omega)\}|=1-\frac{\tilde{q}_{0i}^2}{\overline{q}_{i}^2}$;
\item [ii)] the function $\frac{1}{|\tilde{q}_{0i}^2-q^2-\frac{\epsilon^2}{\tilde{c}_{i}^2}q^2\Delta \tilde{\Sigma}_{i}^1(\textbf{q},\omega)|}$ monotonically decreases by increasing $q$ for $q>\overline{q}_{i}$.
\item [iii)] it is $\overline{q}_{i} \leq q_{Max}^{i}$;
\item [iv)] for $q>\overline{q}_{Max}^{i}$, where $\overline{q}_{Max}^{i}:\frac{\epsilon^2}{\tilde{c}_{i}^2}|Re\{ \Delta \tilde{\Sigma}_{i}^1(\overline{\textbf{q}}^{i}_{Max},\omega)\}|=2[1-\frac{\tilde{q}_{0i}^2}{(\overline{q}_{Max}^{i})^2}]$, it is $\frac{1}{|\tilde{q}_{0i}^2-q^2-\frac{\epsilon^2}{\tilde{c}_{i}^2}q^2\Delta \tilde{\Sigma}_{i}^1(\textbf{q},\omega)|}\leq \frac{1}{|\tilde{q}_{0i}^2-q^2|}$.
\end{enumerate}
The behavior described in points \textit{i)} and \textit{ii)} can be furthrmore observed in Figure 1, {Panels} 3, in the main text. We prove in the following point \textit{iii)}. The other points follow immediately from points \textit{a)-e)}. 
Since $\overline{q}_{i}$ belongs to the \lq high-$q$' region, it is $\tilde{q}_{0i}<\overline{q}_{i}$ and, furthermore, it follows from point \textit{e)}  that $\frac{\epsilon^2}{\tilde{c}_{i}^2}|\Delta \tilde{\Sigma}_{i}^1(\overline{\textbf{q}}_{i},\omega)|\sim \frac{\epsilon^2}{\tilde{c}_{i}^2}|Re\{ \Delta \tilde{\Sigma}_{i}^1(\overline{\textbf{q}}_{i},\omega)\}|=1-\frac{\tilde{q}_{0i}^2}{\overline{q}_{i}^2}\leq 1$ (equal to $1$ in the limit $\frac{\tilde{q}_{0i}}{\overline{q}_{i}} \rightarrow 0$). It is thus $\frac{\epsilon^2}{\tilde{c}_{i}^2} |\Delta \tilde{\Sigma}_{i}^1(\overline{\textbf{q}}_{i},\omega)|\leq1$, from which it follows $\overline{q}_{i}\leq q_{Max}^{i}$.

In the following we define an upper bound for $|\tilde{R}_{k}(\textbf{q},\omega,\epsilon^2)|$.
First, we estimate the contribution to the rest function for $q>q_{Max}^{i*}=max[q_{Max}^{i}, \overline{q}_{Max}^{i}]$. We will call it $\tilde{R}_{k}^*(\textbf{q},\omega,\epsilon^2)=\int_{-1}^{1}dx L_{kkii}(x)\frac{2 \pi}{\tilde{c}_{i}^2}\tilde{r}^*_{i}(\textbf{q},\omega,\epsilon^2,x)$.
From points $\textit{a})$ and $iv)$ it follows that
%\begin{widetext}
\begin{eqnarray}
|\tilde{r}_{i}^*(\textbf{q},\omega,\epsilon^2,x)|=|\int_{q_{Max}^{i*}}^{\infty} dq' c(q,q',x) \frac{{q'}^2}{\tilde{q}_{0 i}^2-q'^2-\frac{\epsilon^2}{\tilde{c}_{i}^2}q'^2 \Delta \tilde{\Sigma}^1_{i}(\textbf{q}',\omega)}| \leq    \int_{q_{Max}^{i *}}^{\infty}dq' c(q,q',x) \frac{{q'}^2}{q'^2-\tilde{q}_{0i}^2}. \label{I_decq_min}
\end{eqnarray}
%\end{widetext}
It is $ \int_{q_{Max}^{i *}}^{\infty}dq' c(q,q',x) \frac{{q'}^2}{q'^2-\tilde{q}_{0i}^2} \sim \frac{1}{\pi^2}\frac{1}{aq_{Max}^{i*}}\leq \frac{1}{a q_{Max}^{i}} $, by recalling that $q_{Max}^{i*} \sim q_{Max}^{i}$, $c(q,q',x)= \frac{a}{\pi^2}\frac{(aq')^2}{(1+(aq')^2+(aq)^2-2(aq)(aq')x)^2}$, $aq_{Max}^{i}\gg 1$ for small values of $\frac{\epsilon^2}{\tilde{c}_i^2}$, and $q, q_{0i}\ll  q_{Max}^{i}$.
Because $\int_{-1}^1dx|L_{kki i}(x)|= O(1)$, finally it is $|\tilde{R}_{k}^*(\textbf{q},\omega,\epsilon^2)| \lesssim \sum_{i} \frac{2}{\pi} \frac{1}{\tilde{c}_{i}^2} \frac{1}{aq_{Max}^{i}}$. 

If $q_{Max}^{i *}=\overline{q}_{Max}^{i}>q_{Max}^{i}$, we need, in addition, to estimate the order of magnitude of 
\begin{eqnarray}
&|\int_{q_{Max}^{i}}^{\overline{q}_{Max}^{i}}dq' c(q,q',x) \frac{{q'}^2}{\tilde{q}_{0i}^2-q'^2-\frac{\epsilon^2}{\tilde{c}_{i}^2}q'^2 \Delta \tilde{\Sigma}_{i}^1(\textbf{q}',\omega)}|. \label{I_qMax}
\end{eqnarray}
We instead take into account the following integral 
\begin{eqnarray}
&\tilde{r}^{\delta}_{i}(x,\textbf{q},\omega,\epsilon^2)=\int_{\overline{q}_{i}-\delta^{i}}^{\overline{q}_{i}+\delta^{i}}dq'c(q,q',x) \frac{{q'}^2}{\tilde{q}_{0i}^2-q'^2-\frac{\epsilon^2}{\tilde{c}_{i}^2}q'^2 \Delta \tilde{\Sigma}_{i}^1(\textbf{q}',\omega)},\nonumber \\ \label{aroundqbar}
\end{eqnarray} 
where $\delta^{i}=\overline{q}_{Max}^{i}-\overline{q}_{i}$. It is $[q^i_{Max},\overline{q}^i_{Max}] \subset [\overline{q}_i-\delta^i,\overline{q}_i+\delta^i]$ because $\overline{q}_{i} \leq q_{Max}^{i}$. We show in the following that $O(|\tilde{r}^{\delta}_{i}(x,\textbf{q},\omega,\epsilon^2)|)=O(|\tilde{r}_{i}^*(x,\textbf{q},\omega,\epsilon^2)|)$. Since the power series expansion of $<G(\textbf{q},\omega)>^1$ defined in Theorem I in the main text converges a.e. for $q<\overline{q}_{i}-\delta^{i}<q_{Max}^{i}$, we can refix the upper integration boundary of the integral in Corrolary II (Eq. 34 in the main text) to $\overline{q}_{i}-\delta^{i}$. This will not affect the domain of the $(\omega,q)$ plane where the GBA can be applied because the contribution to the integral defining the self-energy for $q' \in [\overline{q}_i-\delta^i, \overline{q}_i+\delta^i=\overline{q}^i_{Max}]$ results to be of the same order of magnitude of the remainder function.
The function  $\frac{1}{|\tilde{q}_{0i}^2-q^2-\frac{\epsilon^2}{\tilde{c}_{i}^2}q^2\Delta \tilde{\Sigma}_{i}^1(\textbf{q},\omega)|}$ in the \lq high-$q$' region has a local maximum in $\overline{q}_{i}$, as described in point \textit{i)}, see also Figure 1, {Panels} 3, in the main text. This function is a peak-like function centered in $\overline{q}_{i}$. 
We notice that $\tilde{q}_{0i}^2-\overline{q}_{i}^2-\frac{\epsilon^2}{\tilde{c}_{i}^2}\overline{q}_{i}^2Re\{\Delta \tilde{\Sigma}_{i}^1(\overline{q}_{i},\omega)\} =0$. In a neighbor of $\overline{q}_{i}$ small enough we can thus make the following approximation $1-\frac{\epsilon^2}{\tilde{c}_{i}^2}\frac{q^2}{\tilde{q}_{0i}^2-q^2}Re\{\Delta \tilde{\Sigma}_{i}^1(\textbf{q},\omega)\} \sim A_{i}(q-\overline{q}_{i})$, where $A_{i}=\frac{d}{dq}[1-\frac{\epsilon^2}{\tilde{c}_{i}^2}\frac{q^2}{\tilde{q}_{0i}^2-q^2}Re\{\Delta \tilde{\Sigma}_{i}^1(\textbf{q},\omega)\}]|_{q=\overline{q}_{i}}$. An approximate expression for the constant $A_{i}$ can be obtained by considering that in the \lq high-$q$' region $Re\{\Delta \tilde{\Sigma}_{i}^1(\textbf{q},\omega)\} \sim - C_{i}q^2$, as it can be derived from points \textit{b)} and \textit{c)}. The constant $C_{i}$ should satisfy the condition 
\begin{eqnarray}
&1-\frac{\epsilon^2}{\tilde{c}_{i}^2}\frac{\overline{q}_{i}^2}{\tilde{q}_{0i}^2-\overline{q}_{i}^2}Re\{\Delta \tilde{\Sigma}_{i}^1(\overline{q}_{i},\omega)\} \sim 1-\frac{\epsilon^2}{\tilde{c}_{i}^2} C_{i}\overline{q}_{i}^2 =0, \nonumber
\end{eqnarray}
being $\tilde{q}_{0i}\ll  \overline{q}_{i}$. It is thus $A_{i} \sim -\frac{\epsilon^2}{\tilde{c}_{i}^2}C_{i}2\overline{q}_{i}=- \frac{2}{\overline{q}_{i}}$. We define $\eta_{i} = -\frac{\epsilon^2}{\tilde{c}_{i}^2}\frac{\overline{q}_{i}^2}{\tilde{q}_{0i}^2-\overline{q}_{i}^2} Im\{\Delta \tilde{\Sigma}_{i}^1(\overline{q}_{i},\omega) \} \sim \frac{\epsilon^2}{\tilde{c}_{i}^2} Im\{\Delta \tilde{\Sigma}_{i}^1(\overline{q}_{i},\omega) \}>0$, see point \textit{d)}. Furthermore from point \textit{e)} it follows that $\eta_i\ll  1$.
In the interval $[\overline{q}_{i}-\delta^{i},\overline{q}_{i}+\delta^{i}]$ we can thus take
\begin{eqnarray}
& \frac{1}{\tilde{q}_{0i}^2-q^2-\frac{\epsilon^2}{\tilde{c}_{i}^2}q^2\Delta \tilde{\Sigma}_{i}^1(\textbf{q},\omega)} \sim \frac{1}{\tilde{q}_{0i}^2-q^2} \frac{1}{A_{i}(q-\overline{q}_{i})+i\eta_{i}}.
\end{eqnarray}
It follows that
%\begin{widetext}
\begin{multline}
|\tilde{r}_{i}^{\delta}(\textbf{q},\omega,\epsilon^2,x)| \sim \big| \int_{\overline{q}_{i}-\delta^{i}}^{\overline{q}_{i}+\delta^{i}}dq' c(q,q',x) \frac{{q'}^2}{\tilde{q}_{0i}^2-q'^2} \frac{A_{i}(q'-\overline{q}_{i})}{[A_{i}(q'-\overline{q}_{i})]^2+\eta_{i}^2}  -i \int_{\overline{q}_{i}-\delta^{i}}^{\overline{q}_{i}+\delta^{i}}dq' c(q,q',x) \frac{{q'}^2}{\tilde{q}_{0i}^2-q'^2}\frac{\eta_{i}}{[A_{i}(q'-\overline{q}_{i})]^2+\eta_{i}^2} \big| \sim \\ \frac{\pi}{|A_{i}|} c(q,\overline{q}_{i},x) \frac{{\overline{q}_{i}}^2}{\overline{q}_{i}^2-\tilde{q}_{0i}^2} \sim \frac{1}{2\pi \ a\overline{q}_{i}}. \label{aroundqbar}
\end{multline} 
%\end{widetext}
We assumed that in the integration interval $c(q,q',x)\frac{{q'}^2}{q'^2-\tilde{q}_{0i}^2}\sim c(q,\overline{q}_{i},x)\frac{{\overline{q}_{i}}^2}{\overline{q}_{i}^2-\tilde{q}_{0i}^2} \sim \frac{a}{\pi^2}\frac{1}{(a\overline{q}_{i})^2}$ since $\overline{q}_{i}$ belongs to the \lq high-$q$' region and consequently $a\overline{q}_{i}\gg 1$, $q\ll  \overline{q}_{i}$, $q_{0i}\ll  \overline{q}_{i}$. Furthermore, we observe that the integrand of the first integral in Eq. \ref{aroundqbar} is symmetric with respect to the center of the integration interval. This integral is thus zero. The integrand of the second integral is a Lorentz function of area $\frac{\pi}{|A_{i}|}$. We finally considered that $\eta_{i}\ll  \delta^{i}$, as follows from point \textit{e)}. Because both $\overline{q}_{i}$ and $q_{Max}^{i}$ belong to the \lq high-$q$' region we can finally assume $\frac{1}{a\overline{q}_{i}} \sim \frac{1}{aq^{i}_{Max}}$. 
\section{Supplementary Note 2}
We provide a numerical estimation of the absolute value of the remainder function related to the GBA for given values of frequency and wavevector. We furthermore verify by a numerical estimation that the absolute value of the term $F_k^1(\textbf{q},\omega)$ is significantly larger than such a value. Finally we numerically verify the consistency of the approximation $\Delta \Sigma^1(\textbf{q},\omega) \approx \Delta \Sigma^1(0,\omega)$ while calculating $F_k^1(\textbf{q},\omega)$. To this aim we numerically computed the following integrals, with
$aq_{0L}=1.3$ and $aq=1.2$,
%\begin{widetext}
\begin{eqnarray}
&&\tilde{\Sigma}_{LL}(\textbf{q},\omega)_{<q_{Max}^L}=\int_{-1}^1dxL_{LL}(x) \frac{2 \pi}{\tilde{c}_L^2} \int_{0}^{q_{Max}^L}dq' \ {q'}^2 c(q,q',x)\frac{1}{\tilde{q}_{0L}^2-q'^2-\frac{\epsilon^2}{\tilde{c}_L^2}q'^2\Delta \tilde{\Sigma}_L^{1}(\textbf{q}',\omega)}; \nonumber \\
&&\tilde{R}_{LL}(\textbf{q},\omega,\epsilon^2)=\int_{-1}^1dxL_{LL}(x)\frac{2 \pi}{\tilde{c}_L^2} \int_{q_{Max}^L}^{\infty}dq' \ {q'}^2c(q,q',x) \frac{1}{\tilde{q}_{0L}^2-q'^2-\frac{\epsilon^2}{\tilde{c}_L^2}q'^2\Delta\tilde{\Sigma}_L^{1}(\textbf{q}',\omega)}. \nonumber 
\label{int_num2}
\end{eqnarray}
%\end{widetext}
The theory's input parameters are the same listed in the main text. For such input parameters it is $aq_{Max}^L=18$, as it is possible to observe in Fig. 1, {Panel} 2 -a) in the main text. We obtain $\frac{\mu_0^2}{{c_L^0}^2}|\tilde{\Sigma}_{LL\ <q_{Max}^L}(aq_{0L}=1.3,aq=1.2)|=3.3 \cdot 10^{-1}$ and $\frac{\mu_0^2}{{c_L^0}^2}|\tilde{R}_{LL}(aq_{0L}=1.3,aq=1.2)|=1.0\cdot10^{-3}$. We can compare the latter quantity with the upper bound estimation for $|\tilde{R}_{LL}|$, given in the main text and assessed in Supplemenatry Note 1, i.e. $\sim \frac{\mu_0^2}{{c_L^0}^2} \frac{1}{\tilde{c}_L^2}\frac{2}{\pi} \frac{1}{aq_{Max}^L}=3.0\cdot 10^{-3}$. 
We furthermore numerically evaluate the quantity
\begin{eqnarray}
F^1_{LL}(\textbf{q},\omega)= \int_{-1}^1dxL_{LL}(x) \frac{2 \pi}{\tilde{c}_L^2} \int_{0}^{q_{Max}^L}dq' \ {q'}^2 c(q,q',x)   \frac{\frac{\epsilon^2}{\tilde{c}_L^2}{q'}^2\Delta \tilde{\Sigma}_L(\textbf{q}',\omega)}{(\tilde{q}_{0L}^2-q'^2)^2}, \nonumber
\end{eqnarray}
achieving \footnote{We take the definition $\#\int_a^b \frac{f(x)}{(x-x_0)^2}=lim_{\eta\rightarrow 0}\big[ \int_{a}^{x_0-\eta} \frac{f(x)}{(x-x_0)^2}dx+\int_{x_0+\eta}^{b}\frac{f(x)}{(x-x_0)^2}dx-\frac{2f(x_0)}{\eta}\big]$. It is $lim_{\eta \rightarrow 0} \big[ \int_{\tilde{q}_{0i}-\eta}^{\tilde{q}_{0i}+\eta} q'^2c(q,q',x)\frac{q'^2\Delta \tilde{\Sigma}_{i}(\textbf{q}',\omega)}{(q'^2-\tilde{q}_{0i}^2)^2}dq'-\tilde{q}_{0i}^2c(q,\tilde{q}_0,x)\frac{\Delta \tilde{\Sigma}_{i}(\tilde{q}_{0i},\omega)}{2\eta}\big]=0$. Indeed, given the continuity of the function $c(q,q',x)$ and $\Delta \tilde{\Sigma}_{i}(\textbf{q}',\omega)$ in $\tilde{q}_{0i}$, the quantity in the brackets can be approximated by the expression $\xi(\eta)=c(q,\tilde{q}_0,x) \tilde{q}_{0i}^2\Delta \tilde{\Sigma}_{i}(\tilde{q}_{0i},\omega)\big[ \int_{\tilde{q}_{0i}-\eta}^{\tilde{q}_{0i}+\eta}\frac{q'^2}{(\tilde{q}_{0i}^2-q'^2)^2}$ $dq'-\frac{1}{2\eta}\big]=lim_{\eta \rightarrow 0} \big[\frac{1}{2\eta}(1-\frac{4\tilde{q}_{0i}^2-2\eta^2}{4\tilde{q}_{0i}^2-\eta^2})-ln(\frac{2\tilde{q}_{0i}-\eta}{2\tilde{q}_{0i}+\eta})\big]$. It is hence $lim_{\eta \rightarrow 0} \xi(\eta)=0$.
In the numerical computation we assume $\# \int_{0}^{q_{Max}} q'^2c(q,q',x)\frac{\frac{\epsilon^2}{\tilde{c}_{i}^2}q'^2\Delta \tilde{\Sigma}_{i}(\textbf{q}',\omega)}{(\tilde{q}_{0i}^2-q'^2)^2}dq'=\int_0^{\tilde{q}_{0i}-0.5}q'^2c(q,q',x)\frac{\frac{\epsilon^2}{\tilde{c}_{i}^2}q'^2\Delta \tilde{\Sigma}_{i}(\textbf{q}',\omega)}{(\tilde{q}_{0i}^2-q'^2)^2}dq'+\int_{\tilde{q}_{0i}-0.5}^{q_{Max}}q'^2c(q,q',x)\frac{\frac{\epsilon^2}{\tilde{c}^2}q^2\Delta \tilde{\Sigma}_{i}(\textbf{q}',\omega)}{(\tilde{q}_{0i}^2-q'^2)^2}dq'$. We can take the quantity  $\xi(\eta=0.5)=0.005$ as the related error.} $\frac{\mu_0^2}{{c_L^0}^2}|F^1_{LL}(aq_{0L}=1.3, aq=1.2)|=6.6 \cdot 10^{-2}$, which is a value significantly larger than $|\tilde{R}_{LL}(aq_{0L}=1.3, aq=1.2)|$.

We finally numerically calculate $\frac{\mu_0^2}{{c_L^0}^2}|F_{LL}^{1*}(aq_{0L}=1.3, aq=1.2)|=9.0 \cdot 10^{-2}$, where $F_{LL}^{1*}$ is equal to $F_{LL}^1$ but $\Delta \tilde{\Sigma}_{L}^{1}(\textbf{q},\omega)$ is replaced by $\Delta \tilde{\Sigma}_{L}^1(0,\omega)$. This latter can be compared with the numerical evaluation of $\frac{\mu_0^2}{{c_L^0}^2}|F_{LL}^1(aq_{0L}=1.3, aq=1.2)|$ obtained above, i.e. $6.6 \cdot 10^{-2}$.

\section{Supplementary Note 3}
We show that when in the domain of validity of the GBA the approximation $\Delta \tilde{\Sigma}_{i i}^1(\textbf{q}',\omega) \sim \Delta \tilde{\Sigma}_{i i}^1(0,\omega)$ holds, it is possible to extend the upper integration boundary of the integral defining $F_k^1(\textbf{q},\omega)$ to infinity. The related error is of the same order of magnitude of $|\tilde{R}_{k}(\textbf{q},\omega,\epsilon^2)|$.
We observe that 
\begin{eqnarray}
 \int_{q_{Max}^{i}}^{\infty}dq' \ {q'}^2 c(q,q',x) \frac{q'^2}{(q'^2-\tilde{q}_{0i}^2)^2}|\frac{\epsilon^2}{\tilde{c}_{i}^2}\Delta \tilde{\Sigma}_{i i}^1(0,\omega)|\ \leq \int_{q_{Max}^{i}}^{\infty}dq' \ {q'}^2 c(q,q',x)\frac{q'^2}{(q'^2-\tilde{q}_{0i}^2)^2} \sim \frac{1}{aq_{Max}^{i}}.
\end{eqnarray}
In the region of frequency where $\frac{\epsilon^2}{\tilde{c}_{i}^2}\Delta \tilde{\Sigma}^{1,Max}_{i}(\omega)<1$ indeed it is $\frac{\epsilon^2}{\tilde{c}_{i}^2}|\Delta \tilde{\Sigma}_{i i}(0,\omega)|<1$. Furthermore for $q'>q_{Max}^{i}$, it is $\tilde{q}_{0i}\ll  q'$.
If the approximation  $\Delta \tilde{\Sigma}_{i i}^1(\textbf{q}',\omega) \sim \Delta \tilde{\Sigma}_{i i}^1(0,\omega)$ holds, the Hadamard Principal value of the integral defining $F_k^1(\textbf{q},\omega)$ can be calculated by exploiting the Residue Theorem because the function $z^2c(q,z,x)\frac{[z^2\frac{\epsilon^2}{\tilde{c}_{i}^2}\Delta \tilde{\Sigma}_{i i}^1(0,\omega)]}{(z^2-\tilde{q}_{0i}^2)^{2}}$ has only non-essential singularities in the complex plane.

\section{Supplementary Note 4}
We show that as long as the condition $|\frac{\Delta \tilde{\Sigma}_{i i}^1(\textbf{q}',\omega)-\Delta \tilde{\Sigma}_{i i}^1(0,\omega)}{\Delta \tilde{\Sigma}_{i i}^1(0,\omega)}|<\frac{1}{2}$ is fulfilled the dominant contribution to the integral defing $F_k^1(\textbf{q},\omega)$ can be obtained trough the approximation $\Delta \tilde{\Sigma}_{i i}^1(\textbf{q}',\omega) \sim \Delta \tilde{\Sigma}_{i i}^1(0,\omega)$.
It is
\begin{multline}
F_k^1(\textbf{q},\omega)=lim_{\eta \rightarrow 0^+}\int_{-1}^1dxL_{kk i i}(x) \frac{2 \pi}{\tilde{c}_{i}^2} \int_{0}^{q_{Max}^{i}}dq' \ {q'}^2 c(q,q',x)  \big\{\frac{\frac{\epsilon^2}{\tilde{c}_{i}^2}q'^2\Delta \tilde{\Sigma}_{i i}(0,\omega_{\eta})}{(\tilde{q}_{0i,\eta}^2-q'^2)^2}+ \frac{\frac{\epsilon^2}{\tilde{c}^2}q'^2[\Delta \tilde{\Sigma}_{i i}(\textbf{q}',\omega_{\eta})-\Delta \tilde{\Sigma}_{i i}(0,\omega_{\eta})]}{(\tilde{q}_{0i,\eta}^2-q'^2)^2}\big\}.
\label{I1_0}
\end{multline}
If $|\frac{\Delta \tilde{\Sigma}_{i i}^1(\textbf{q},\omega)-\Delta \tilde{\Sigma}_{i i}^1(0,\omega)}{\Delta \tilde{\Sigma}_{i i}^1(0,\omega)}|<\frac{1}{2}$ in the integration interval $[0,q_{Max}^{i}]$ the integral of the first term of the summation in Eq. \ref{I1_0} is the dominant. In support of this statement we observe that $lim_{\eta \rightarrow 0^+}\int_{0}^{q_{Max}^{i}}dq' \ q'^2c(q,q',x) \frac{1}{2} \frac{\frac{\epsilon^2}{\tilde{c}^2}q'^2}{|\tilde{q}_{0i,\eta}^2-q'^2|^2} \leq lim_{\eta \rightarrow 0^+}|\int_{0}^{q_{Max}^{i}}dq' \ q'^2 c(q,q',x) \frac{\frac{\epsilon^2}{\tilde{c}^2}q'^2}{(\tilde{q}_{0i,\eta}^2-q'^2)^2}|$. The inequality is obtained by exploiting Eqs. 22 and 24 with $N=1$ in the text, recalling that $ c(q,q',x) \in \mathbb{R}^+$ and $|\mathrm{Re}[z]|+|\mathrm{Im}[z]|\leq 2 |z|$, where z is a complex number.

\end{document}